\documentclass[prb,aps,twocolumn]{revtex4}

\usepackage{times}
\usepackage{graphicx}
\usepackage{float}
\usepackage{latexsym,amsmath,amssymb,bm,euscript}
\usepackage{color}
\usepackage{subfigure}
\usepackage{epstopdf}
\usepackage[colorlinks=true,linkcolor=blue,citecolor=blue]{hyperref}
\usepackage{type1cm}
\fontsize{12pt}{18pt}

\usepackage[titletoc]{appendix}
\usepackage{appendix}

\begin{document}

\title{Emergent Chiral Spin Liquid:  Fractional Quantum Hall Effect in a Kagome Heisenberg Model}
\author{Shou-Shu Gong, Wei Zhu, and D. N. Sheng}
\affiliation{Department of Physics and Astronomy, California State University, Northridge, California 91330, USA.}

\begin{abstract}
The fractional quantum Hall effect (FQHE) realized in two-dimensional electron systems under a magnetic field  is one of the most remarkable discoveries
in condensed matter physics.
Interestingly, it has been proposed that FQHE can also emerge in time-reversal invariant spin systems, known as the chiral spin liquid (CSL) 
characterized
by the topological order and the emerging of the fractionalized quasiparticles. 
A CSL can naturally lead to the exotic superconductivity originating from the condense of anyonic quasiparticles.
Although CSL was highly sought after for more than twenty years,
it had never been found in a spin isotropic  Heisenberg model or related materials.
By developing a  density-matrix renormalization group based method for adiabatically inserting flux, we discover  a FQHE in a spin-$\frac{1}{2}$ isotropic kagome Heisenberg model.
We identify this FQHE state as the long-sought CSL with a uniform chiral order spontaneously breaking time reversal symmetry, 
which is uniquely characterized by  the half-integer quantized topological Chern number protected by a robust excitation gap.
The CSL is found to be at the neighbor of the
previously identified $Z_2$ spin liquid, which may lead to an
exotic quantum phase transition between two gapped topological spin liquids.
\end{abstract}
\maketitle

The experimentally discovered fractional quantum Hall effect (FQHE) \cite{Tsui1982,Laughlin1983,Girvin} is the first 
demonstration of topological order and fractional (anyonic) statistics \cite{Wen1990,Halperin1982,Wen19902,Wen1991,Haldane1983} realized in two-dimensional electronic systems
under a magnetic field breaking time-reversal symmetry (TRS).
A related new state of matter with fractionalized quasiparticle excitations is the topological quantum spin liquid (QSL)
emerging in frustrated magnetic systems \cite{Balents2010,Anderson1987, Kivelson1988,Read,Sondhi2001,Fisher2000,Lesik2002,Balents2002,Kitaev2006_1,Lee2008}.
Such spin systems, related to strongly correlated Mott  materials and holding the clue
to the unconventional superconductivity in doped  systems,  are of fundamental importance
to the condensed matter field\cite{Balents2010,Anderson1987,Lee2006,Lee2008, Xu}.
To understand the emergent physics of  frustrated magnetic systems, where  spins escape from the
conventional fate of developing symmetry broken ordering, the concept of QSL with the fractionalized
quasiparticles was established\cite{Anderson1987,Wen1991,Fisher2000}.
Experimental candidates for such a new state of matter are identified 
including kagome antiferromagnets\cite{Janson2008,Fak2012,YSLee2012}
and triangular organic compounds \cite{Shimizu2003,Kurosaki2005,Itou2008}.
The simplest QSL with TRS is the gapped $Z_2$ spin liquid, which possesses the $Z_2$ topological order
and fractionalized spinon and vison quasiparticle excitations\cite{Wen1991,Fisher2000}.
The $Z_2$ QSL is identified as an example of the resonating valence-bond liquid state, which was first proposed by
Anderson\cite{Anderson1987}.
Although explicitly demonstrated in contrived theoretical systems\cite{Kivelson1988,Sondhi2001,Balents2002,Kitaev2006_1},
the searching of the gapped   QSL in realistic Heisenberg models has always attracted much attention over the last twenty years.
The primary example is the recent discovered gapped $Z_2$ QSL for kagome Heisenberg model (KHM) with the
dominant nearest neighbor (NN) interactions based on the density-matrix renormalization group (DMRG)
simulations\cite{White2011, Jiang2012, Depenbrock2012, Jiang2008}.

Another class of QSL with fractionalized quasiparticles obeying fractional (anyonic) statistics  is chiral spin liquid (CSL)
\cite{Laughlin1988,Wen1989,Wilczek,Haldane1995,KYang}, 
which breaks TRS and parity symmetry while preserves other lattice and spin rotational symmetries.
Kalmeyer and Laughlin\cite{Laughlin1988} first proposed that, in a time-reversal invariant spin system with geometry frustration,
one can realize a $\nu=1/2$ FQHE  as a CSL state\cite{Wen1989}  through mapping the frustrated in-plane 
exchange interactions to the uniform magnetic field.
A CSL is also considered to be a simple way in which frustrated spin systems develop topological
order through statistics transformation to cancel out the frustration\cite{Wen1989, Wilczek}.
The CSL may also lead to the exotic anyon superconductivity with doping holes into such systems\cite{Wen1989,Wilczek}.
The existence of CSL through spontaneously TRS breaking has been demonstrated in a
Kitaev model on a decorated honeycomb lattice with contrived anisotropic spin interactions\cite{Yao2007}
and most recently in a spin anisotropic kagome model\cite{YCHe}.
Interestingly, based on the classical and Schwinger boson mean-field analyses,
QSLs with different chiral spin orders 
have been suggested for extended KHM \cite{Lhuillier2012,Lhuillier2013}.
Other theoretical studies show that one can also induce a CSL state through adding multi-spin TRS breaking chiral 
interactions\cite{Greiter2007,Thomale2009, Cirac2012,Bauer2013}.
Although CSL has been explored for more than twenty years \cite{Laughlin1988, Wen1989, Wilczek, Haldane1995, KYang, Bauer2013, Hermele2009, Lhuillier2012},
the accurate DMRG\cite{White2011,Jiang2012,Depenbrock2012,Jiang2008} and variational Monte Carlo\cite{Yasir2014} studies on various frustrated Heisenberg
models often lead to the conventional ordered phases or TRS preserving $Z_2$ and $U(1)$ QSLs.
The simple concept of realizing CSLs of the nature of FQHE through spontaneously breaking TRS and statistics transformation\cite{Wen1989,Wilczek}
remains  illusive in realistic frustrated magnetic systems.

\begin{figure}
   \centering
   \includegraphics[width=1.0\linewidth]{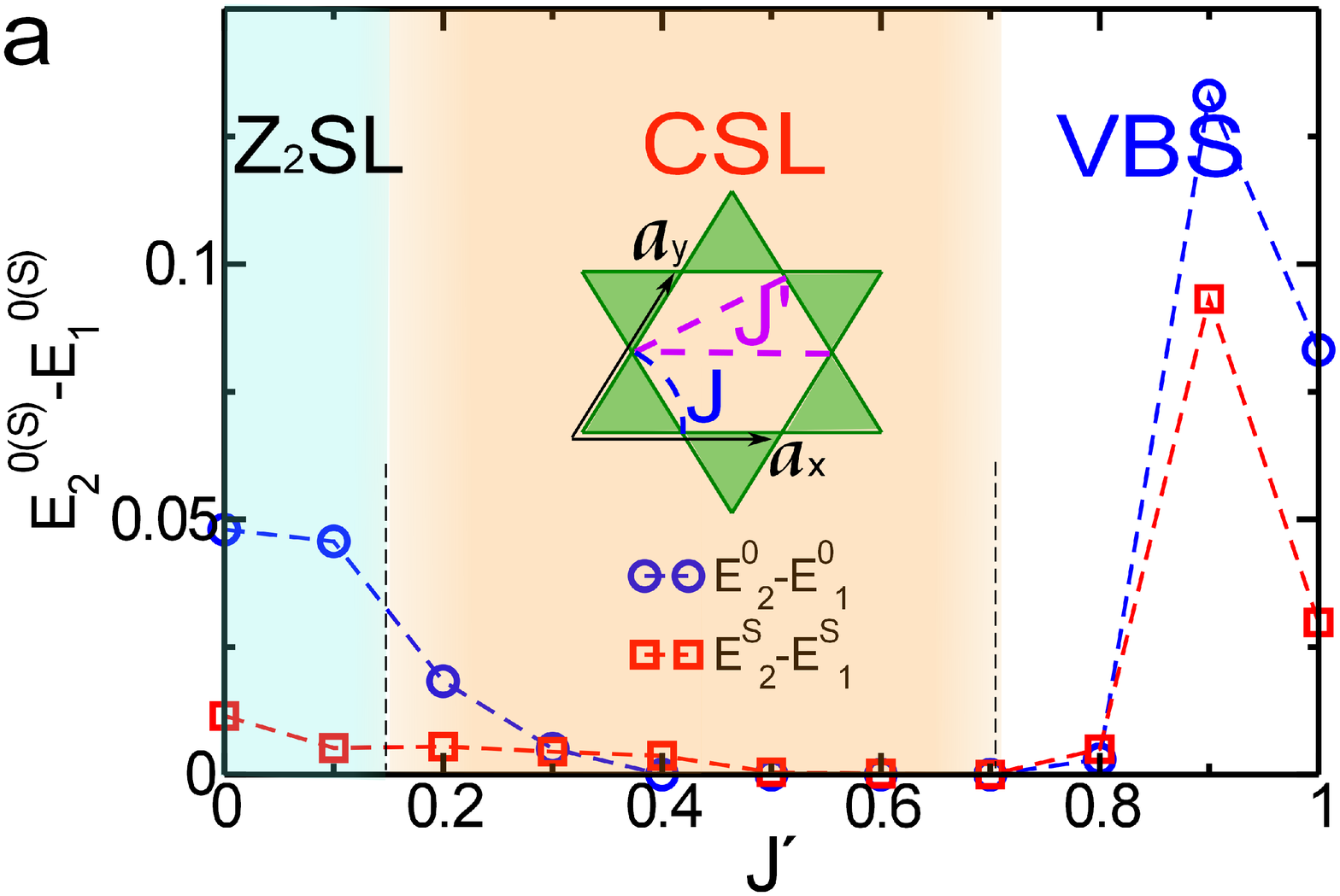}
   \includegraphics[width=1.0\linewidth]{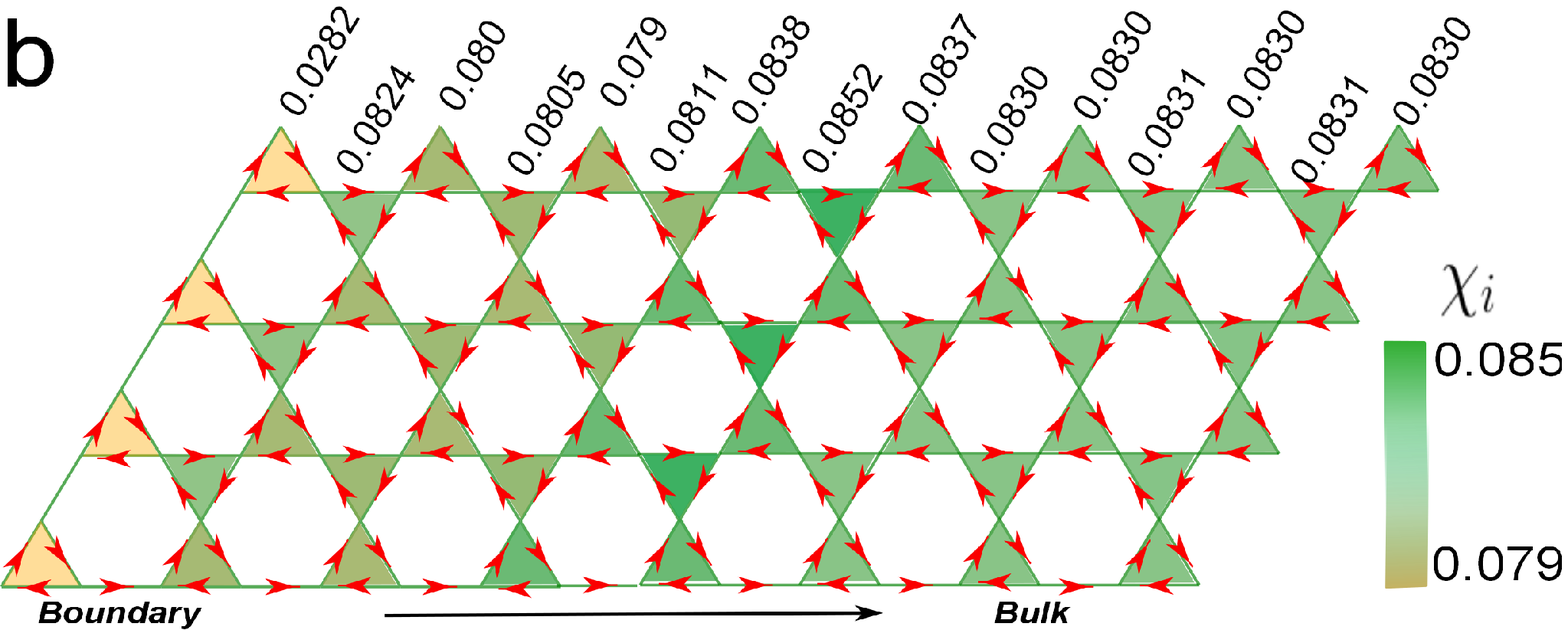}
   \caption{
\textbf{Phase diagram of the spin-$\frac{1}{2}$ $J-J^{\prime}$  antiferromagnetic kagome Heisenberg model (KHM)}.
We set the NN coupling $J=1$, and the second and third NN couplings as $J^{\prime}$.
In DMRG simulations, we study cylinders with
lengths $L_x$ and $L_y$ (in unit cells) along the $x$ and $y$ (the tilt lattice vector) directions, respectively.
We also label our system by the total number of sites $N=3\times L_x\times L_y$.
(a) The energy differences between the two lowest states $E^{0}_2-E^{0}_1$ ($E^S_2-E^S_1$) in two different topological 
(vacuum and S-) sectors are shown for a $N=3\times 24\times 4$ cylinder at different $J^{\prime}$, which 
reveal the four degenerating ground states in two  sectors
for  a range of parameter $0.2\lesssim J^{\prime} \lesssim 0.7$.
(b) We illustrate the uniform positive chiral order  $\langle \chi_i \rangle = \langle S_{i_1}\cdot(S_{i_2}\times S_{i_3})\rangle$ $(i_1,i_2,i_3
\in \bigtriangleup_i (\bigtriangledown_i))$ on a $N=3\times 24\times 4$ cylinder with $J^{\prime}=0.5$
measured from the MES $|{\tilde \psi^{0}_{L}}\rangle$ in the vacuum sector (identified in Fig.~\ref{entropy}), which breaks 
the TRS and parity symmetry.
The CSL phase is characterized by the long-range chiral correlations and a fractionally quantized $C=1/2$ Chern number,
which identifies the state as the Laughlin $\nu=1/2$ FQHE emerging in the $J-J^{\prime}$ KHM.
}\label{phase}
\end{figure}

In this article, we report a new theoretical discovery of the CSL in an extended spin-$\frac{1}{2}$ KHM based on the state of art DMRG simulations\cite{White1992,Schollwock2005}. As illustrated in the inset of Fig.~\ref{phase}(a), the system has the NN coupling $J=1$ as energy scale, as well as the second and third NN couplings 
$J^{\prime}$ inside each hexagon of the kagome lattice, described by the following Hamiltonian \cite{Balents2002,Lhuillier2012}:
\begin{equation}
H = J \sum_{\langle i,j\rangle}S_i\cdot S_j + J^{\prime} \sum_{\langle\langle i,j\rangle\rangle}S_i \cdot S_j + J^{\prime} \sum_{\langle\langle\langle i,j\rangle\rangle\rangle}S_i \cdot S_j.
\end{equation}
We perform the numerical flux insertion simulations on cylinder systems based on the newly developed adiabatical DMRG to detect
the topological Chern number, which uniquely characterizes the chiral spin liquid.
We have fully established a robust $\nu = 1/2$ FQHE state for $0.1 \lesssim J^{\prime} \lesssim 0.7$ by observing the half-integer quantized topological Chern number protected by
a robust excitation gap, the degenerate ground states, and the uniform chiral order spontaneously breaking TRS.\\

\noindent \textbf{Results}

\noindent \textbf{Phase diagram.} Our main findings are summarized in the phase diagram Fig.~\ref{phase}(a).
With the turn on of a positive $J^{\prime}$, we find a robust CSL phase in the region
of $0.1 \lesssim J^{\prime} \lesssim 0.7$.
We design and perform the Laughlin flux insertion numerical experiment through developing an adiabatic DMRG, which
inserts flux and obtains the ground state for each flux.
The adiabatic DMRG allows us to obtain the  topological Chern number\cite{Girvin,Haldane1995}, which characterizes the topological
nature of the quantum phase.  Our simulation experiment shows that the CSL is characterized by a fractionally quantized
Chern number $C=1/2$, which is a ``smoking gun'' evidence of the emergent
$\nu=1/2$  Laughlin FQHE state\cite{Laughlin1988}  in the frustrated KHM.
The CSL phase is also characterized by a four-fold degeneracy in two topological sectors.
In each sector, there is a double degeneracy representing the two sets of CSL states with opposite chiralities.
The near uniform chiral order measured for a state spontaneously breaking TRS is illustrated in Fig.~\ref{phase}(b). 
We also establish that the CSL is neighboring with the $Z_2$ QSL previously found\cite{White2011,Jiang2012,Depenbrock2012,Jiang2008} at $J^{\prime}=0$, while the transition region 
appears to be under strong influence of the nonuniform Berry curvature resulting from gauge field, 
which may provide new insights to many puzzles regarding theoretical\cite{White2011,Jiang2012,Depenbrock2012,Jiang2008}
and experimental findings\cite{Lee2008,Janson2008,Fak2012,YSLee2012} for kagome antiferromagnets.\\

\begin{figure}
   \centering
   \includegraphics[width=1.0\linewidth]{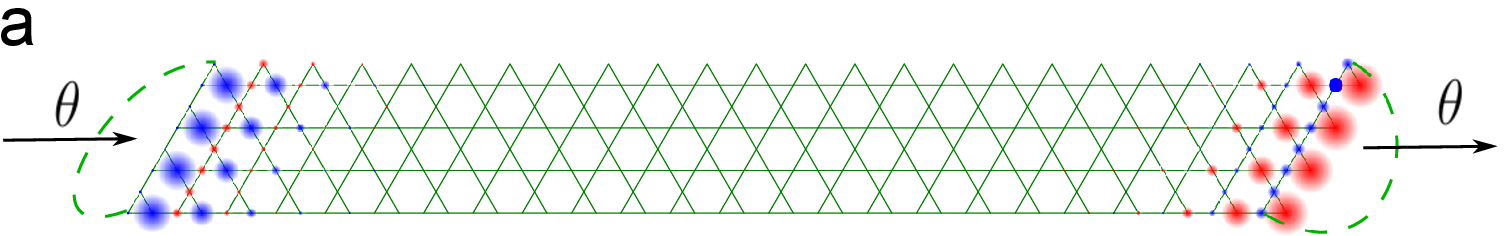}
   \includegraphics[width=0.9\linewidth]{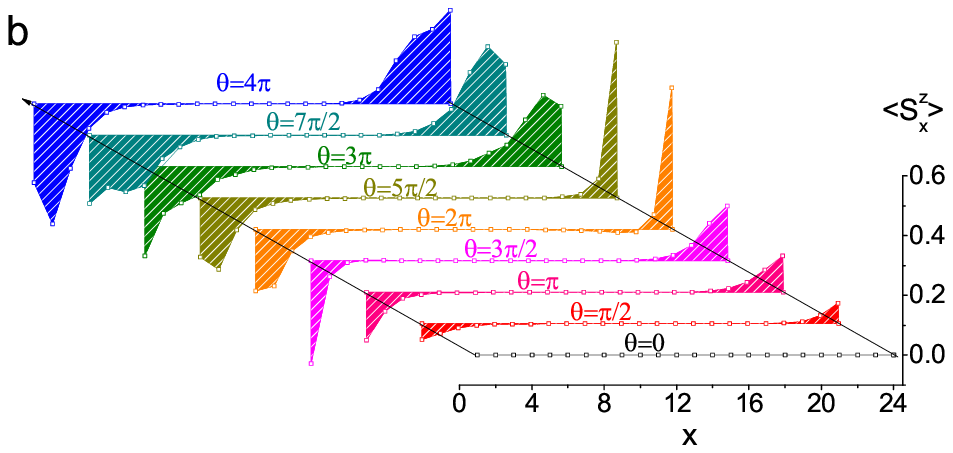}
   \includegraphics[width=0.7\linewidth]{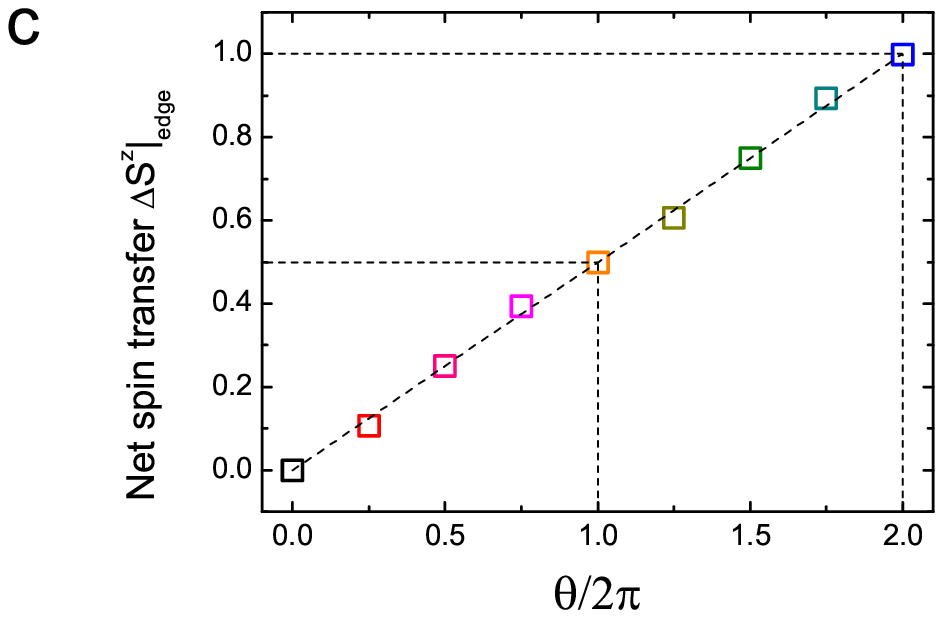}
   \caption{\textbf{Laughlin flux insertion Gedanken experiment and fractionalized Chern number $C=1/2$ for CSL.}
   (a) Real-space configuration of the spin magnetization $\langle S^z_{x,y}\rangle$ at position $R_i=(x,y)$
after adiabatically inserting a quantized  flux $\theta=2\pi$. The area of the circle is proportional to the amplitude of $\langle S^z_{x,y}\rangle$. The red (blue) color represents the positive (negative) $\langle S^z_{x,y}\rangle$.
   (b) Real-space configuration of the accumulated spin magnetization $\langle S^z_x\rangle = \sum_{y}\langle S^z_{x,y}\rangle$ 
(the summation is over all the $3L_y$ sites in each column $x$) with increasing flux $\theta$.
Clearly, we see  a net spin-z accumulating in the right edge of the sample,
which is equivalent to the transfer of hardcore bosons (the hardcore boson number  $n_i$ is related to the on-site $S^z_i$ as $n_i=S^z_i+1/2$)
being pumped from the left edge to the right edge without going through the bulk.
So this simulation experiment reveals a quantum Hall system with a nonzero Hall conductance, while the bulk is gapped.
(c) Net spin transfer  $ \Delta S^z|_{\rm edge}$ to the right edge of the cylinder as a
function of  $\theta$.
From the net spin transfer in one period of flux $\theta= 0 \rightarrow 2\pi$, we obtain the
exact fractionally quantized Chern number $C=\Delta S^z|_{\rm edge}=1/2$.
      The results are demonstrated for a $3\times 24\times 4$ cylinder at $J^{\prime}=0.5$ using the $U(1)$ DMRG with keeping up to $5000$ states.
Similar results are obtained for all the states within the CSL phase.
   }\label{chern}
\end{figure}

\noindent \textbf{Fractional quantization of topological number.} 
To uncover the full topological nature of the  phase at large system scale, we perform
the flux inserting simulation based on the adiabatic DMRG.   
For conventional FQHE systems,  a quantized net charge transfer would appear as $\Delta N =C$ from one edge
of the sample to the other edge after inserting one period of flux $\theta=0\rightarrow 2\pi$,
corresponding to a nonzero fractionally quantized topological invariant Chern number $C$\cite{Haldane1995}, which is
$C=1/2$ for the $\nu=1/2$ bosonic Laughlin state.

By adiabatically inserting the flux $\theta$ in our DMRG experiment,
we study the evolution of the local magnetization $\langle S^z_{x,y}\rangle$, which is the spin-z average of the ground state
at a local lattice site $R_i=(x,y)$.
With the increase of $\theta$, we
measure the corresponding spin accumulations of each ground state at $\theta=j\pi/2$ ($j$ is an integer).
One example with $\theta=2\pi$ is shown in Fig.~\ref{chern}(a).
We find nonzero magnetization
starting to  build up at the left and right edges of cylinder,
which grows monotonically with the growing of $\theta$ as shown in Fig.~\ref{chern}(b).
Since our system has total spin conservation,  the net spin-z transfer $\Delta S^{z}|_{\rm edge}$ (which is the total magnetization around the right edge of the system) is equivalent to the pumping of the hardcore bosons from the left edge to the right edge without going through the bulk.
In Fig.~\ref{chern}(c),  we show the net spin transfer $\Delta S^z|_{\rm edge}$ as a function of $\theta$.
A near linear spin pump is being realized in this chiral spin state, which is exactly quantized
as $\Delta S^z|_{\rm edge}=0.5$ at $\theta=2\pi$.  From the fundamental correspondence between
edge spin transfer and bulk Chern number\cite{Hatsugai1993}, we identify the bulk Chern number of
the system as $C=1/2$, fully characterizing the state as the Kalmeyer-Laughlin CSL\cite{Laughlin1988} of $\nu=1/2$ FQHE.
Physically, the pumping in FQHE system is achieved through the adiabatical rotation of the basis states
of the many-body wavefunction, which can be viewed as a non-local operation by developing a ``spinon'' line in the cylinder.
We find the entanglement spectrum of the spinon sector obtained here by inserting $2\pi$ flux is identical to the one of the S-sector shown below in Fig.~\ref{entropy}(b) obtained through pinning.
With further increasing the flux to $\theta = 4\pi$, the net spin transfer 
$\Delta S^z|_{\rm edge} = 1.0$, where  the system evolves back to the vacuum sector.
These observations fully establish the bosonic $\nu = 1/2$ FQHE emerging in the $J-J^{\prime}$ KHM.
While the Chern number simulations characterize  the ground state
as the long-sought CSL, we will further measure the topological degeneracy, chiral correlations,
topological entanglement entropy, and modular matrix to demonstrate the full nature of the topological 
state in our time-reversal invariant system.\\

\begin{figure}
   \centering
   \includegraphics[width=0.48\linewidth]{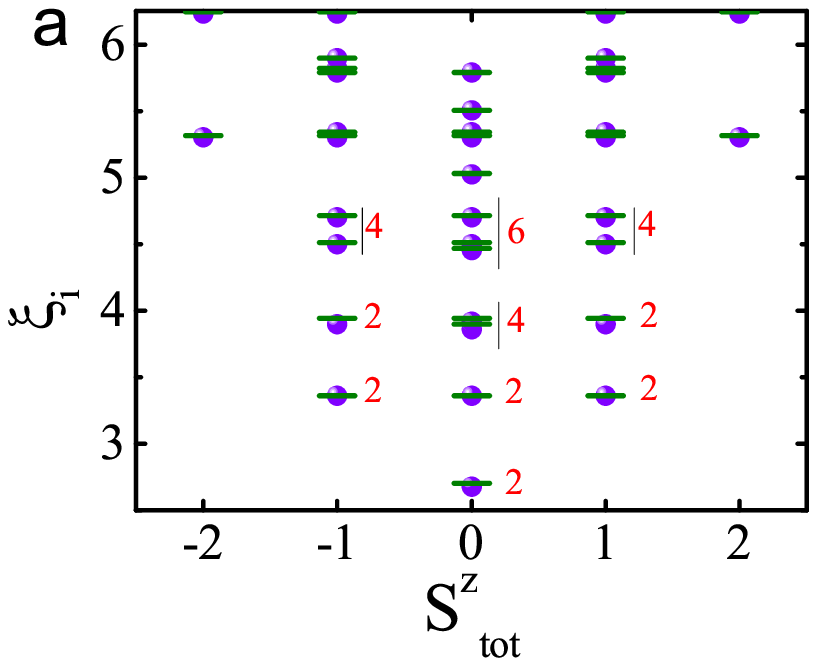}
   \includegraphics[width=0.49\linewidth]{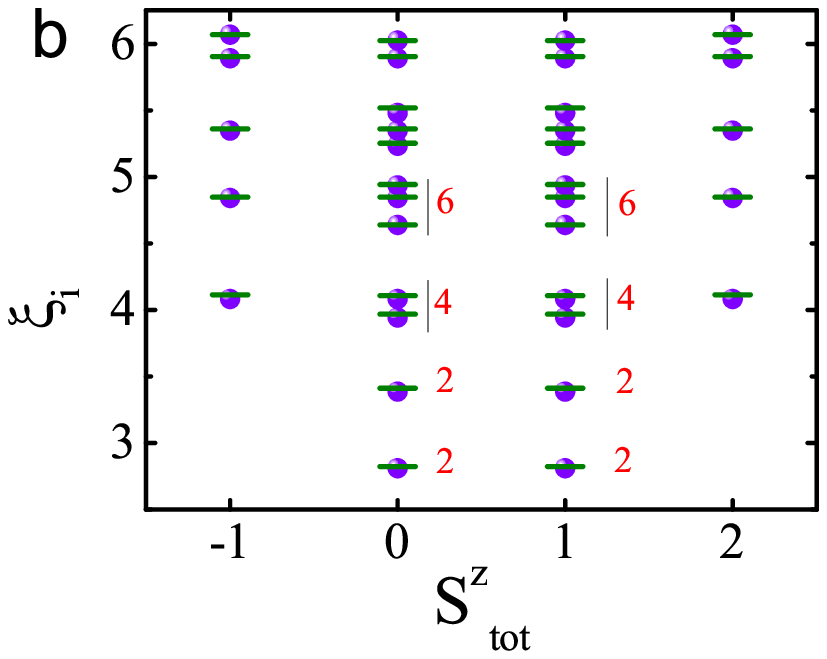}
   \includegraphics[width=0.6\linewidth]{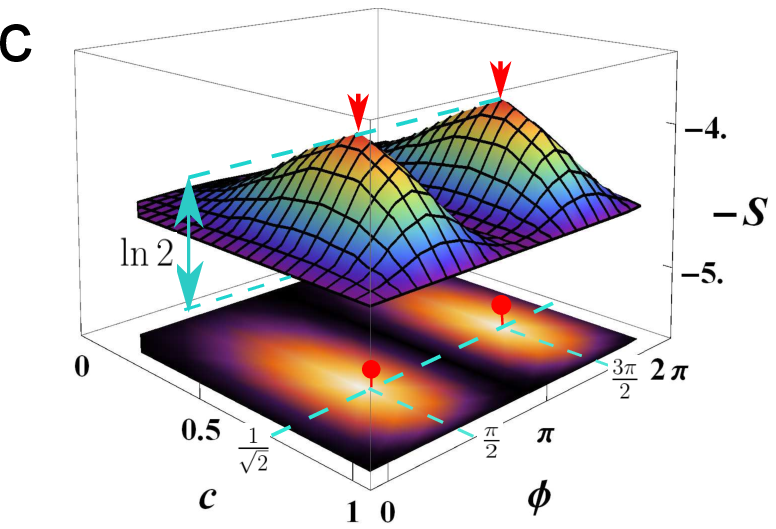}
   \caption{\textbf{Entanglement spectra characterization and MESs for the TRS broken phase}. 
(a) and (b) are the entanglement
  spectra of the ground states in the vacuum ($|\psi^{0}_{1}\rangle$) and S- ($|\psi^{S}_{1}\rangle$)
  sectors, respectively. $\xi_i =-\ln \lambda_i$ with $\lambda_i$ the eigenvalues of reduced density matrix.
The two degenerate states $|\psi^{0}_{1,2}\rangle$ ($|\psi^{S}_{1,2}\rangle$) have exactly the same spectra.
These states are real wavefunctions consistent with the TRS of the model Hamiltonian.
  The lines and circles with the near identical $\xi_i$ denote the double degeneracy indicating that each of these low-energy states is a maximum entropy state.
The numbers ${2,2,4,6}$ label the near degenerating pattern for the low-lying entanglement spectra
which are doubled from what one would expect for a Laughlin FQHE state from conformal field theory.
   (c) Entanglement entropy profile $-S$ of the general  superposition state  in the vacuum sector
  $|{\tilde \psi^{0}}\rangle  = c|\psi^{0}_1\rangle + \sqrt{1-c^2}e^{i\phi}|\psi^{0}_2\rangle$.
  The MESs that are pointed by red arrows and dots emerge as $|{\tilde \psi^{0}_{L,R}}\rangle  = \frac{1}{\sqrt{2}} (|\psi^{0}_1\rangle \pm i|\psi^{0}_2\rangle)$.
The MES $|{\tilde \psi}^0_{R(L)}\rangle$ has uniform counterclockwise (clockwise) chiral order for each triangle as illustrated in Fig.~\ref{phase}(b).
Furthermore, if we initiate the DMRG state as a complex state, we automatically find such a MES,
which spontaneously breaks TRS.
The results are demonstrated for $J^{\prime}=0.5$ for a $N=3\times 24\times 6$ cylinder, and near identical
results are obtained for different parameters  in the CSL phase.
    }\label{entropy}
\end{figure}

\noindent \textbf{Low-energy spectrum and topological degeneracy.} 
The Kalmeyer-Laughlin CSL has two-fold topological ground-state degeneracy, and the spontaneously TRS breaking for such a time-reversal invariant system must have an additional double degeneracy in each topological sector.
On cylinder geometry, one can control the boundary condition
near the cylinder edges to target into different topological sectors\cite{White2011, YCHe2013}, which we denote as the vacuum
and S-sectors, respectively.
By using this technique in DMRG,
we find the two lowest-energy states in each sector
whose energy differences $E^0_2-E^0_1$ and $E^S_2-E^S_1$
drop to small values for $0.2 \lesssim J^{\prime} \lesssim 0.7$.
One example is shown in Fig.~\ref{phase}(a) for a cylinder with $L_x=24$ and  $L_y=4$.
Importantly,  the degenerating states $|\psi^0_{1,2}\rangle$ ($|\psi^S_{1,2}\rangle $)
in each topological sector also have  near identical entanglement spectra.
The double degeneracy of entanglement spectrum for the ground states
$|\psi^{0(S)}_1\rangle$ is explicitly shown using
two different symbols (line and circle) in Figs.~\ref{entropy}(a) and \ref{entropy}(b).
These observations are consistent with the spontaneously TRS breaking double degeneracy.
We also find the ground-state energies between the two sectors are degenerate ($E^S_1/N-E^0_1/N=0.00001$ for $J^{\prime}=0.5$ at $L_y=4$), which, combined with the distinct entanglement spectra \cite{Haldane2008} as shown in Figs.~\ref{entropy}(a) and \ref{entropy}(b) of the two sectors, 
establish the topological degeneracy for these two sectors in the intermediate phase.
By searching for other low energy excited states from both DMRG and exact diagonalization (ED),  we exclude that there are other distinct topological
degenerating sectors for the intermediate region, while a lot more lower energy states appear near $J^{\prime}=0$.

The energy and entanglement spectra doubling are  signatures of finding the maximally
entangled states  in each sector, which is forced by the TRS
of the system Hamiltonian (here we used a real number initial wavefunction in DMRG calculations
which forbids any spontaneous TRS breaking).
To demonstrate the nature of the new quantum phase,
we first find the minimum entangled states (MESs) in each topological sector \cite{YZhang2012, Jiang2012, WZhu2013},
which represent the eigenstates of the Wilson-loop (string-like) operators encircling the cylinder
and are the simplest states of the quasiparticles.
In Fig.~\ref{entropy}(c), we show two MESs emerging (labeled by two red dots) in the vacuum sector:
$|{\tilde \psi^0_{L(R)}}\rangle=\frac{1}{\sqrt 2} (|\psi^{0}_1\rangle \pm i|\psi^{0}_2\rangle)$, which are equal magnitude
superposition of the real states with a phase difference $\pm \pi/2$.
The MES $|{\tilde \psi^0_L}\rangle$ breaks the TRS spontaneously and
demonstrates a uniform nonzero chirality order for each triangle
as illustrated  in Fig.~\ref{phase}(b). 
The chiral order reaches a value around $0.08$ comparable to its classical value $1/8$.
The conjugate state $|{\tilde \psi^0_R}\rangle$ as another MES has the opposite sign of chirality.
The doubling of the entanglement spectra for the maximum entropy state simply results from the superposition of
the MESs with the same entanglement spectra. 
Consequently, one finds an entanglement entropy
difference $\ln 2$ comparing to the MESs as illustrated in  Fig.~\ref{entropy}(c).
Near identical results and two MESs are also found in the topological degenerating S-sector.
Furthermore, if we initiate the DMRG state with  a random complex number state, we automatically find such a MES,
which spontaneously breaks TRS.

By obtaining the MES, we find the topological entanglement entropy $\gamma$ consistent with the result $\ln 2 / 2$ of
the $\nu = 1/2$ Laughlin state\cite{Kitaev2006,Levin2006}. The ED calculations further confirm  this state on a $N=3\times 4\times 3$ cluster by extracting modular transformation matrix 
\cite{YZhang2012,WZhu2013}  from the MESs of two noncontractable 
cuts (see Supplementary Information for more details). 
\\

\begin{figure}
   \centering
   \includegraphics[width=1.0\linewidth]{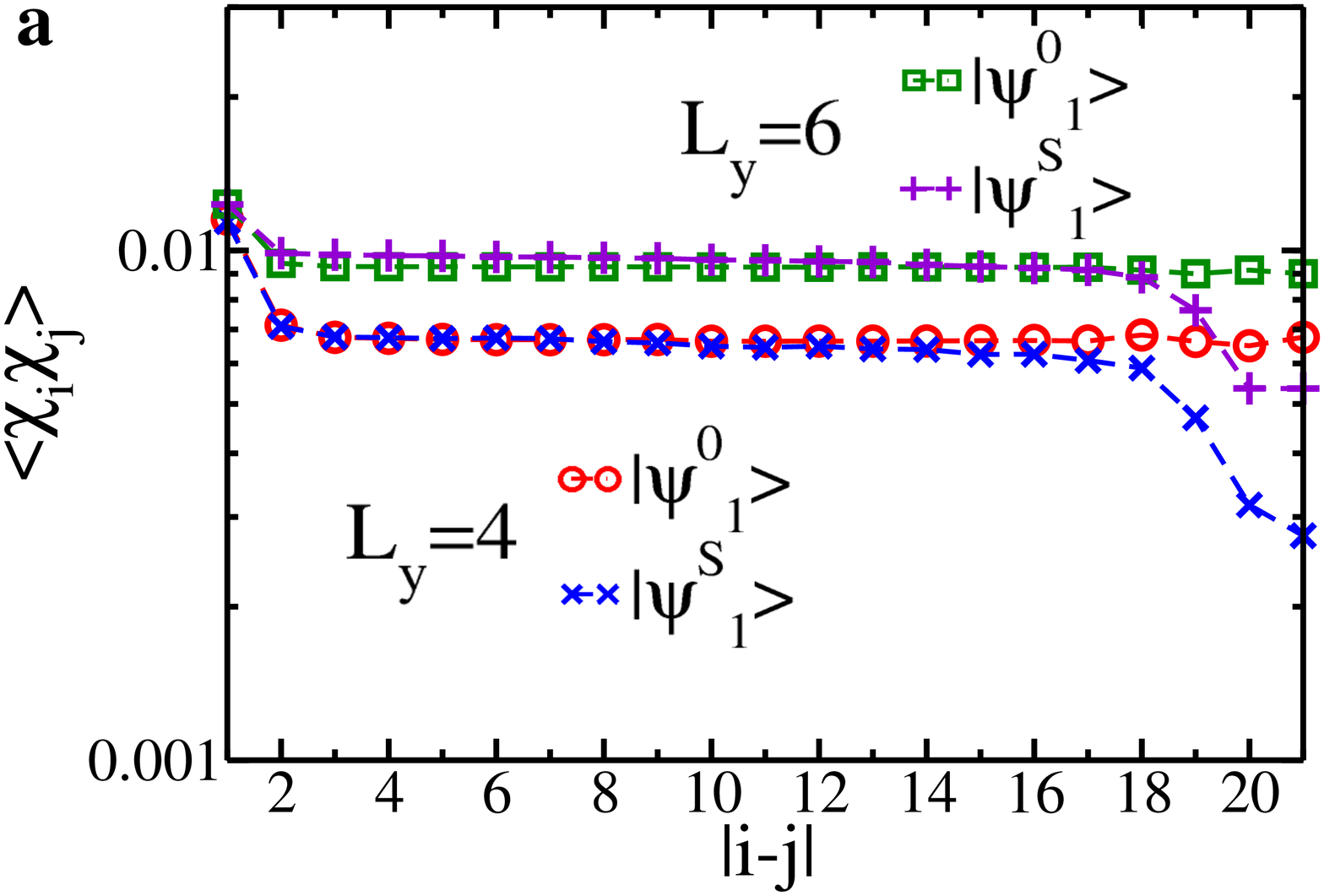}
   \includegraphics[width=1.0\linewidth]{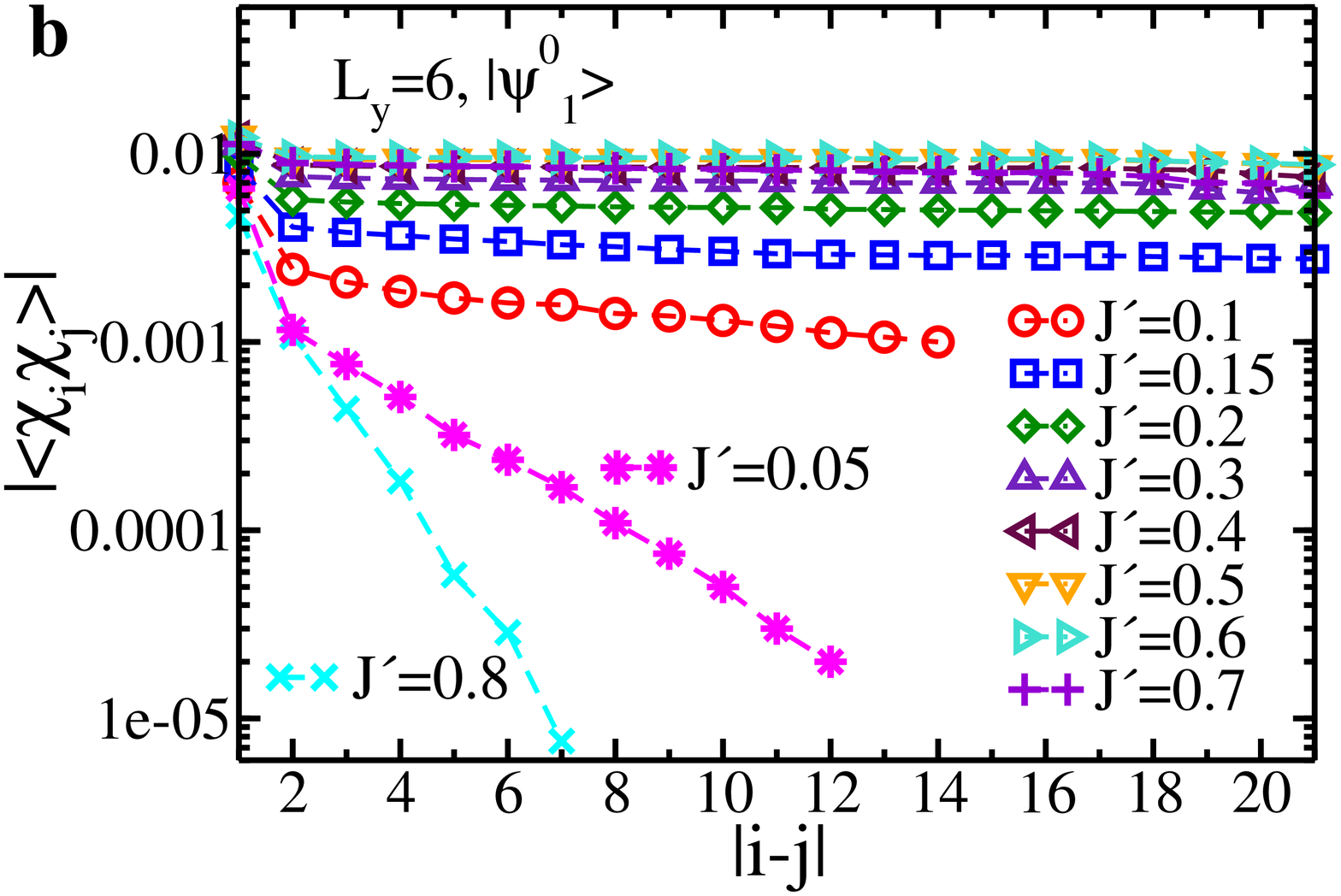}
   \caption{\textbf{Long-range chiral-chiral correlation function for CSL}.
  (a) Log-linear plot of the  chiral-chiral correlation function $\langle \chi_i\chi_j\rangle$  versus the
   distance of triangles $|i-j|$ along the $x$ direction on the $N=3\times 24\times 4$ and $3\times 24\times 6$ cylinders
   for the ground states in both vacuum and S-sectors at $J^{\prime} = 0.5$. All the correlations are
positive and increase with the system width $L_y$. In the S-sector, $\langle \chi_i\chi_j\rangle$ have small edge effects
due to the presence of two localized spins from pinning.
(b) Log-linear plot of absolute correlation $|\langle \chi_i\chi_j\rangle|$
  versus $|i-j|$ along the $x$ direction for $N=3\times 24\times 6$ cylinders at various $J^{\prime}$ obtained from the ground state in the vacuum sector.
  All the correlations demonstrate long-range order (and they are also positive) for the intermediate phase
$0.1 \leq J^{\prime} \leq 0.7$, while they decay exponentially at $J^{\prime}=0.05$ and $0.8$, 
where transitions to the $Z_2$ spin liquid and VBS phase take place. 
The correlations are near a constant everywhere in the CSL phase, so we choose $i=3$ and vary $j$ to the end of the cylinder.
}\label{chiral}
\end{figure}

\noindent \textbf{Quantum phase transitions.}
We use both the chiral-chiral correlation functions and  the topological Chern number obtained from inserting flux to  identify the
quantum phase diagram and transitions in the $J-J^{\prime}$ model.
In Fig.~\ref{chiral}(a), we compare the chiral correlations $\langle\chi_i\chi_j\rangle$ for the states from the two
topological sectors with different system widths at $J^{\prime}=0.5$.
We find long-range correlations for the states from both topological sectors,
which are further enhanced  with increasing system width $L_y$.
To reveal the quantum phase transitions, we
show the chiral correlation functions calculated from the ground state of the vacuum sector for different $J^{\prime}$ in Fig.~\ref{chiral}(b).
$\langle \chi_i\chi_j\rangle$ is positive everywhere and has the long-range order for $0.1 \leq J^{\prime} \leq 0.7$,
while transitions to other phases are detected at $J^{\prime}=0.05$ and $0.8$ by identifying the exponential decaying chiral correlations.

In the flux insertion simulations, we find that the Chern number remains to be quantized at
$C=1/2$ for the same parameter range $0.1 \leq J^{\prime} \leq 0.7$, thus we establish the quantum phase diagram as shown in Fig.~\ref{phase}(a). The quantum phase transition around $J^{\prime} \sim 0.7-0.8$ is charaterized by an excitation gap closing in the bulk of system, where we detect a strong bulk magnetization (boson density) response to the inserted flux.
Between $J^{\prime}=0$ and $0.1$,  we detect a strong nonuniform Berry curvature resulting from the gauge field in the inserting flux simulations, possibly indicating the forming of new quasiparticles and the emerging of $Z_2$ QSL.
We also study the stability of the CSL when the second and third neighbor couplings are different. We find the CSL phase in a  region around the line with $J_2 = J_3$. For example, when $J_2 = 0.1$, the CSL is robust for $0.1 \lesssim J_3 \lesssim 0.3$.
Physically, the $J_3$ coupling  suppresses the magnetic order formed in the $J_1-J_2$ ($J_2 \simeq 0.2$) kagome model, thus
substantially enlarges the non-magnetic region. Meanwhile, classically the $J_3$ term will enhance a noncoplanar 
spin chiral order\cite{Lhuillier2012},  which may induce a CSL in the quantum $J_1-J_2-J_3$ model as demonstrated here.\\

\noindent \textbf{Discussion}

\noindent In the past twenty  years,  the gapped QSL in realistic magnetic systems have attracted intensive attention. While the NN
or $J_1-J_2$ KHM \cite{White2011, Jiang2012, Depenbrock2012, Jiang2008} is the primary candidate of a possible $Z_2$ QSL, there are still many puzzles left unresolved.
The frustrated kagome antiferromagnets Herbertsmithite Cu$_3$(Zn,Mg)(OH)$_6$Cl$_2$ and Kapellasite Cu$_3$Zn(OH)$_6$Cl$_2$\cite{Lee2008,Janson2008,Fak2012,YSLee2012}
are possible candidates of QSL; however, they appear to be more consistent with gapless or critical states.
At theoretical side, redundant low-energy excitations are found for the NN KHM from ED simulations \cite{Lauchli2011},
variational studies find that  $U(1)$ gapless QSL \cite{Ran2007,Yasir2014} has relatively low energy,
and DMRG studies have not been able to identify all the four topological sectors for $Z_2$ QSL \cite{YCHe2013}.
Our finding of the robust CSL at the neighbor of the NN KHM indicates that the latter is not a fully developed $Z_2$ QSL yet,
and the nature of states for the experimental relevant kagome systems may be strongly affected by a new  quantum critical point between two
gapped QSLs, the $Z_2$ and the CSL.
In a parallel work, a CSL has also been uncovered in an anisotropic kagome spin system\cite{YCHe} with only
spin-z interactions for further neighbors. 
We believe that our numerical findings will stimulate new theoretical and experimental researches in this field to resolve the nature of the quantum phases for different frustrated magnetic systems. An exciting next step will be identifying theoretical models and experimental materials  which can host exotic topological superconductivity by doping different CSLs.\\


\noindent \textbf{Methods}

\noindent
DMRG is a powerful tool to study the low-lying states of strongly correlated electron systems\cite{White1992}.
The accuracy of DMRG is well controlled by the number of kept states $M$, which denotes the $M$ eigenstates
of the reduced density matrix with the  largest eigenvalues.
The highly efficiency of DMRG for one-dimensional systems or two dimensional 
cylinder systems have been shown for different systems \cite{Schollwock2005, White2011}.
An improvement in DMRG calculations is to implement symmetry to reduce the Hilbert space.
The spin-z  or total particle $U(1)$ symmetry is commonly used in DMRG, which is preserved in many
model systems. 
For some systems with spin rotational $SU(2)$ symmetry such as the Heisenberg spin model,
the more efficient choice is to apply the $SU(2)$ symmetry \cite{SU2}, from which we can obtain more accurate results
for wider systems. This algorithm has been applied to study various  frustrated Heisenberg systems successfully \cite{Depenbrock2012,ssgong2013,ssgong2014}.\\

\noindent \textbf{Details of the $SU(2)$ DMRG calculation.} We study the frustrated KHM without flux using $SU(2)$ DMRG.
We study the cylinder system with open boundaries in the $x$ direction and periodic boundary condition in the $y$ direction.
For $L_y=4$ ($L_y=6$) systems, we keep up to $3000$ ($4600$) $SU(2)$ states with the DMRG truncation error
$\epsilon \simeq 1\times 10^{-6}$ ($\epsilon \simeq 1\times 10^{-5}$) for most calculations.
To find the ground states in both vacuum and S- topological sectors on cylinders in the DMRG calculations,
we take pinning sites in the open boundaries or insert flux to target the two different sectors \cite{YCHe2013}.\\

\noindent \textbf{Adiabatic DMRG and fractionally quantized Chern number.} For the first time, we develop the numerical flux insertion experiment for cylinder systems based
on the adiabatical DMRG simulation  to  detect the topological Chern number\cite{Haldane1995} of the bulk system, which uniquely characterizes the CSL as a $\nu=1/2$ FQHE state emergent from the $J-J^{\prime}$  Heisenberg model on kagome lattice.
In this simulation, we  impose the twist  boundary conditions along the $y$ direction by replacing terms
$S_i^+S_j^-+h.c.  \rightarrow e^{i\theta} S_i^+S_j^-+h.c.$ for all neighboring $(i,j)$ bonds with interactions crossing the $y$-boundary in the Hamiltonian. Starting from a small $\theta\sim 0$,  a state with the definite
chirality  and sign of Chern number will be randomly selected, which remains the same through out the whole
adiabatical process of $\theta =0 \rightarrow 4\pi$. We find states
with the opposite Chern numbers ($C=\pm 1/2$) in different runs of the simulations due to spontaneously TRS breaking.
A robust excitation gap $\Delta \sim 0.24$ is obtained for $J^{\prime}=0.5$ after we create two spinons (at $\theta=2\pi$)
at the opposite edges  of the cylinder (see Fig.~\ref{chern}(a)), which protects the CSL state.
This method can be  applied to study different interacting systems and characterize 
different topological states.



\noindent \textbf{Acknowledgements} We thank Y. C. He for extensive discussions.
We also thank Leon Balents, Matthew P. A. Fisher, Olexei I. Motrunich  and  F. Duncan M. Haldane for stimulating discussions and
explanations of spin liquid as well as topological physics.
This research is supported by the National Science Foundation through grants 
DMR-1205734 (S.S.G.), DMR-0906816 and DMR-1408560 (D.N.S.), the U.S. Department of
Energy, Office of Basic Energy Sciences under grant No.
DE-FG02-06ER46305 (W.Z.).\\

\noindent \textbf{Author Contributions} S.S.G. and W.Z. performed main calculations based on different numerical programs they
developed. S.S.G., W.Z. and D.N.S. made significant contributions from the design of the
project to the finish of the manuscript.\\

\noindent \textbf{Additional information}\\
\textbf{Competing financial interests:} The authors declare no competing financial interests.


\newpage
\begin{appendices}

\begin{center}
\textbf{Supplementary Information}
\end{center}

\section{Topological degenerate ground-state energy}

In the gapped topological states, the ground-state energies in different topological sectors are
near degenerate on finite-size systems. With the increase of the system width, the difference of the near degenerate
energies vanishes exponentially. In the density-matrix renormalization group (DMRG) calculations, we obtain the bulk energy per site in both the vacuum $(E^0_1/N)$
and spinon $(E^S_1/N)$sectors by subtracting the energies of two long cylinders with different system lengths in each sector.
$(E^S_1-E^0_1)/N$ describes the difference of the ground-state energies in different topological sectors.

Here, we show the results for $J^{\prime} = 0.5$ on $L_y = 4$ cylinder in Suppl. Table.~\ref{energy}.
By keeping the unconverged $2000$ $SU(2)$ states, we find a small energy difference $0.00008$. And with increasing kept states,
the difference continues to decrease. For the well converged ground states with the DMRG truncation error $\epsilon \simeq 1\times 10^{-7}$
by keeping $5000$ $SU(2)$ states, we show the energy difference is $0.00001$, which is consistent with the exact diagonalization
results shown below and the topological degeneracy in the system. This energy splitting $0.00001$ is much smaller than that in
the nearest-neighbor (NN) kagome Heisenberg model $0.00069$.

\begin{table}
\begin{tabular}{|c|c|c|c|}
\hline
$J^{\prime}=0.5,L_y=4$ & $E^0_1/N$ & $E^S_1/N$ & $(E^S_1-E^0_1)/N$ \\
\hline
$M_{\rm SU(2)}=2000$ & $-0.46046$ & $-0.46038$ & $0.00008$ \\
\hline
$M_{\rm SU(2)}=4000$ & $-0.46050$ & $-0.46048$ & $0.00002$\\
\hline
$M_{\rm SU(2)}=5000$ & $-0.46052$ & $-0.46051$ & $0.00001$\\
\hline
\end{tabular}
\caption{\textbf{Degenerate ground-state energies in the different topological sectors}. The ground-state energy per site in both the vacuum ($E^0_1/N$) and spinon ($E^S_1/N$) sectors, as well as the energy difference
between the two sectors $(E^S_1-E^0_1)/N$ for $J^{\prime}=0.5$ on the $L_y=4$ cylinder. To avoid edge effects, these bulk energies are obtained by subtracting
the energies of two long cylinders with different system lengths. $M_{\rm SU(2)}$ is the kept $SU(2)$ states.} \label{energy}
\end{table}

\section{Topological entanglement entropy}

For the gapped quantum states with topological order, the topological entanglement entropy (TEE) $\gamma$ is proposed to characterize
the non-local entanglement. 
The Renyi entropy of a subsystem $A$ with reduced density matrix $\rho_A$ are defined as $S_n = (1-n)^{-1}\ln({\rm Tr}\rho_A^{n})$,
where the $n \rightarrow 1$ limit gives the Von Neuman entropy. For a topologically ordered state, Renyi entropy has the form $S_n = \alpha L - \gamma$,
where $L$ is the boundary of the subsystem, and all other terms vanish in the large $L$ limit; $\alpha$ is a non-universal constant, while a positive
$\gamma$ is a correction to the area law of entanglement and reaches a universal value determined by total quantum dimension $D$ of quasiparticle excitations
as $\gamma = \ln D$. For the $\nu = 1/2$ Laughlin state, the quantum dimension of each quasiparticle is $1$, leading to 
the total dimension $D = \sqrt{2}$ and thus the TEE $\gamma = \ln 2 / 2$.

By using the complex number DMRG simulations, we obtain the minimal entropy state (MES) with spontaneously broken time-reversal
symmetry and the corresponding Von Neuman entanglement entropy. With the help of the $SU(2)$ DMRG, we could obtain the converged  entropy 
for $L_y = 3,4,5$ cylinders. For $L_y = 6$ cylinder, we cannot get the converged entropy because the required DMRG optimal state
number $M_{\rm SU(2)}$ is beyond our computation abilities. Thus, to find an estimation of the entropy on $L_y = 6$ cylinder, we study the entropy
versus $1/M_{\rm SU(2)}$ as shown in Fig. \ref{entropy}(a), and make a careful extrapolation of the data to estimate the converged
result. For $J^{\prime} = 0.5$, we find the entropy $S = 4.49 \pm 0.02$. In Fig. \ref{entropy}(b), we make a linear fitting of the
entropy data for $L_y = 4,5,6$ cylinders at $J^{\prime} = 0.5$, and find the TEE $\gamma = 0.34 \pm 0.04$, which is consistent with the TEE of the $\nu = 1/2$
Laughlin state $\gamma = \ln 2 / 2$.

\begin{figure}
   \centering
   \includegraphics[width=1.0\linewidth]{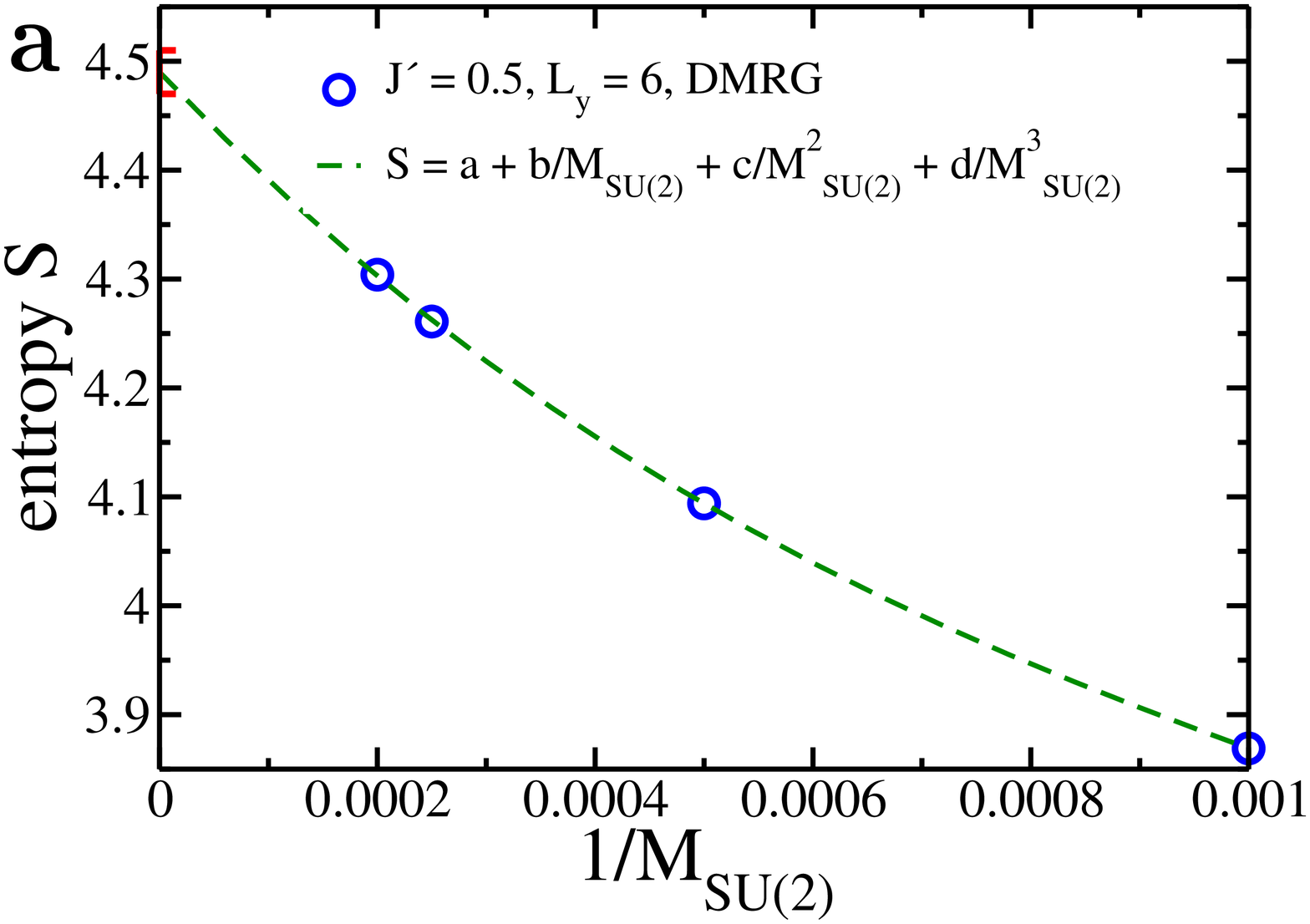}
   \includegraphics[width=1.0\linewidth]{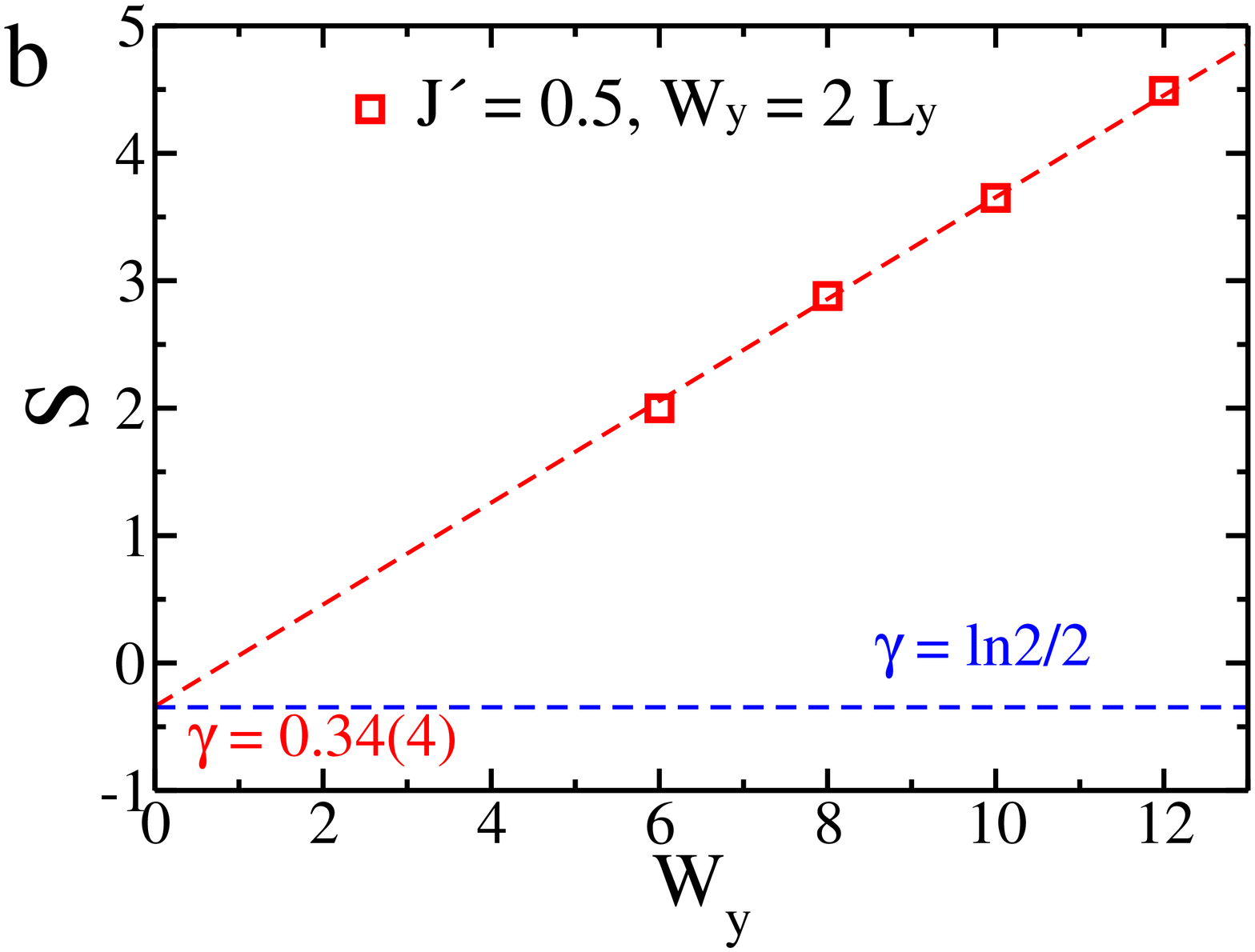}
   \caption{\textbf{Topological entanglement entropy}. (a) Entanglement entropy $S$ versus DMRG optimal state number inverse
$1/M_{\rm SU(2)}$ for $J^{\prime} = 0.5$ on $L_y = 6$ cylinder. The DMRG entropy is for the MES, which is obtained from the 
complex number DMRG simulations. The data are fitted using the formula $S = a + b/M_{\rm SU(2)} + c/M_{\rm SU(2)}^2 + d/M_{\rm SU(2)}^3$,
from which we find the convergent entropy $S = 4.49 \pm 0.02$. (b) Entanglement entropy versus system width for $J^{\prime} = 0.5$.
By a linear fitting of the results for $L_y = 4,5,6$, we find the TEE $\gamma = 0.34 \pm 0.04$, where the error bar is from
the uncertainty of the entropy on $L_y = 6$ cylinder as shown in (a). The TEE we find is consistent with the result of the
$\nu = 1/2$ Laughlin state $\gamma = \ln 2 / 2$.}\label{entropy}
\end{figure}

\section{Spin-Spin correlation function}

\begin{figure}
   \centering
   \includegraphics[width=1.0\linewidth]{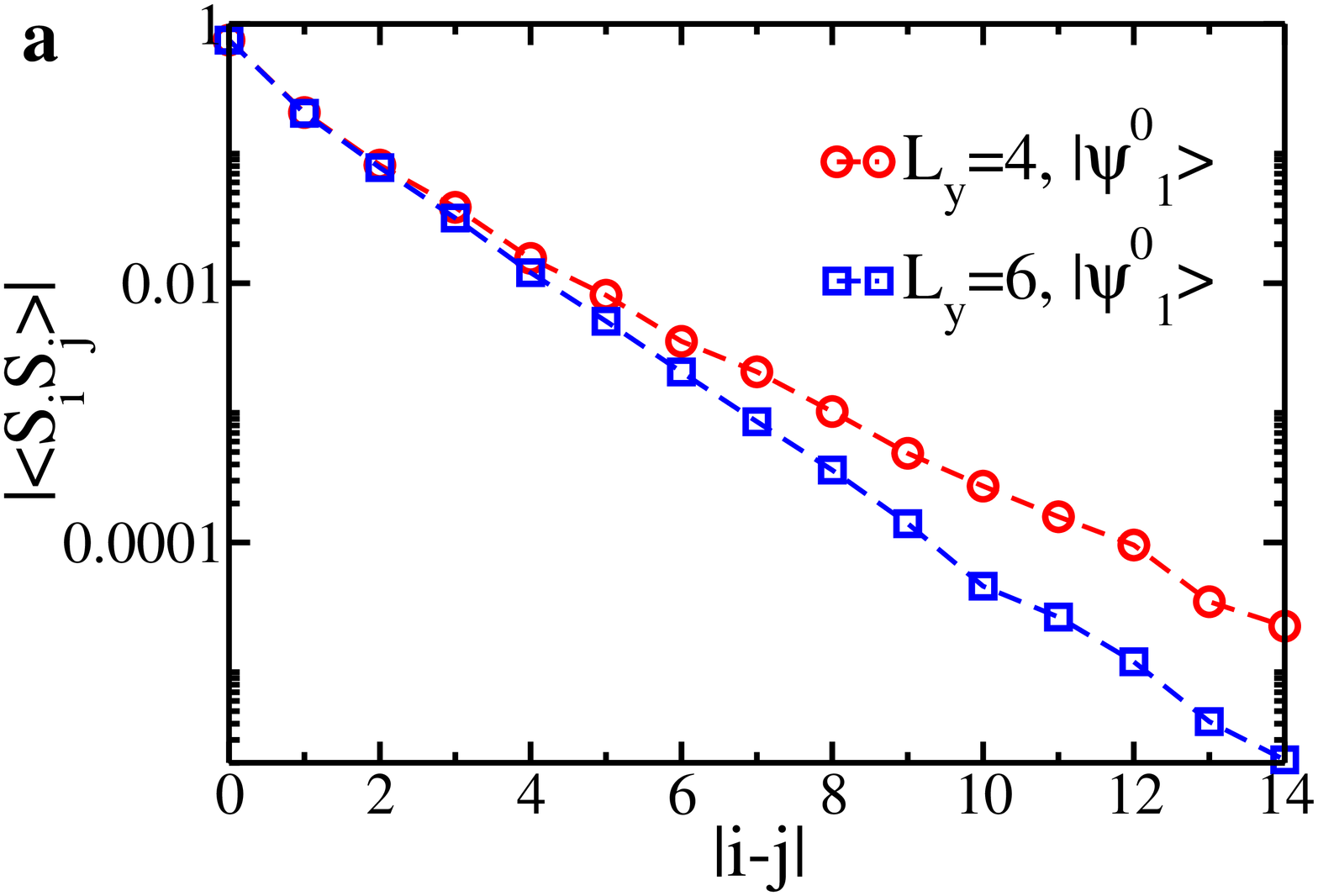}
   \includegraphics[width=1.0\linewidth]{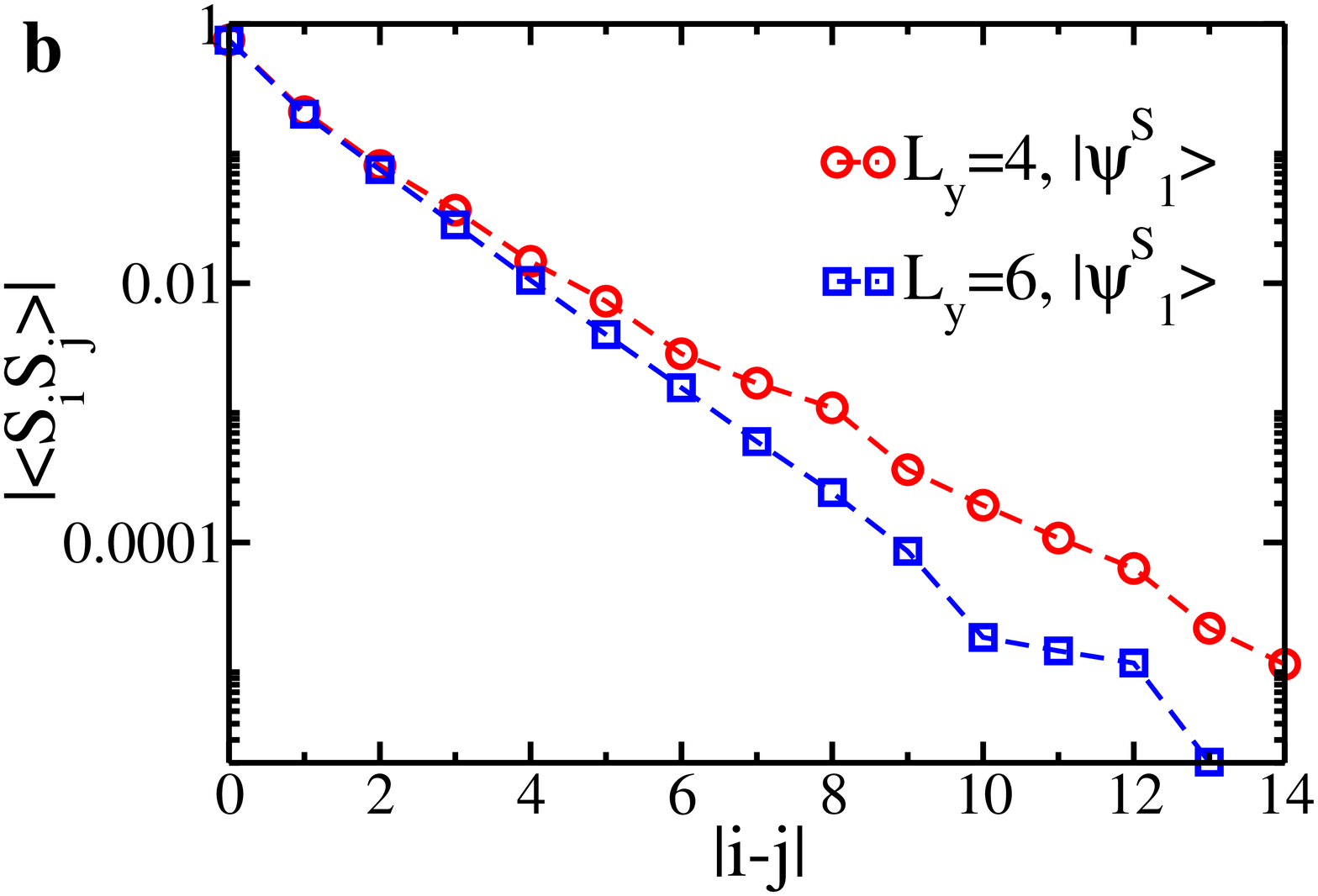}
   \caption{\textbf{Spin-Spin correlation function}. Log-linear plot of the absolute value of the spin-spin correlation
  function versus the distance of sites $|i-j|$ on $3\times 24\times 4$ and $3\times 24\times 6$ cylinders for
  $J^{\prime} = 0.5$ in (a) vacuum and (b) S-sectors. The reference site $i$ is located in the center of cylinder, and
  site $j$ is chosen along the same row from the bulk to the boundary.}\label{spincor}
\end{figure}

For a gapped topological chiral spin liquid (CSL), the system is expected to have a short-range spin correlation.
We measure the spin-spin correlation function on the cylinders with $L_y = 4$ and $6$ for both vacuum and spinon sectors.
We demonstrate $\langle S_i\cdot S_j\rangle$ with site $i$ in the center of lattice and $j$
along the same row from the bulk to the boundary for $J^{\prime} = 0.5$ in Suppl. Fig.~\ref{spincor}.
The spin correlations exhibit the exponential decay in both vacuum and spinon sectors.
And the decay length does not increase with growing $L_y$ from $4$ to $6$, which suggests that
the spin correlation length is close to saturation with growing system width. This observation is consistent
with a vanishing magnetic order.

\section{Valence-bond solid order}

\begin{figure*}
   \centering
   \includegraphics[width=1.0\linewidth]{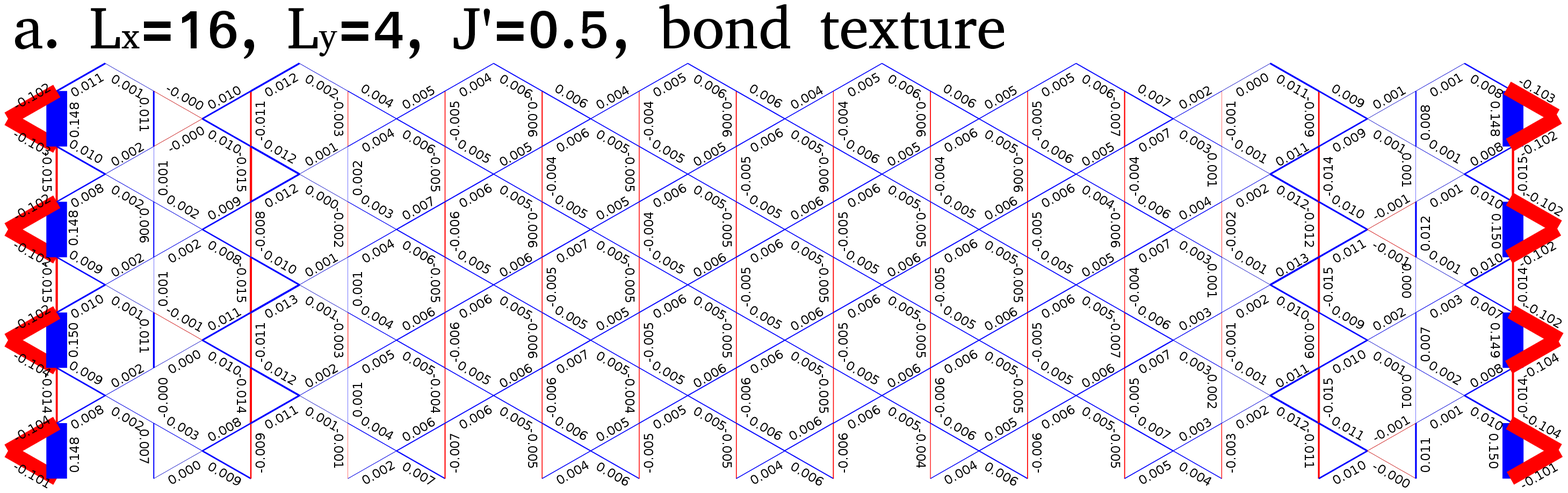}
   \includegraphics[width=1.0\textwidth]{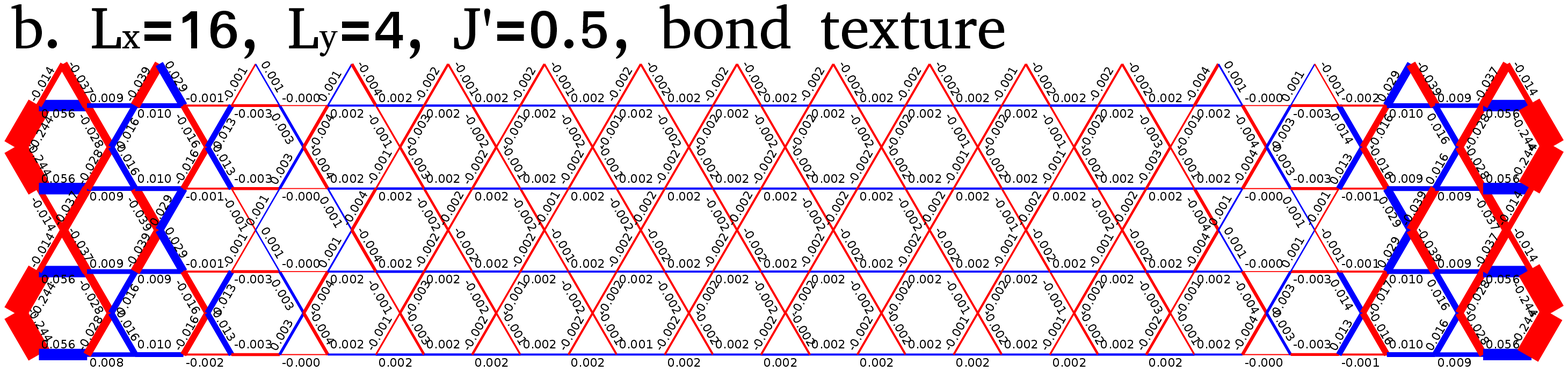}
   \caption{\textbf{Nearest-neighbor bond textures at $J^{\prime}=0.5$}. The NN bond textures $B_{i,j}$ on the $3\times 16\times 4$ cylinders at
  $J^{\prime} = 0.5$ in the vacuum sector. The numbers denote the amplitudes of bond texture $B_{i,j} = \langle S_i \cdot S_j\rangle-e$,
  where $e$ is the average NN bond energy in the bulk of cylinder. Here, we find $e = -0.212$ and $-0.210$ for $(a)$ and $(b)$, respectively.
   The blue (red) color represents the positive (negative) bond texture.}\label{bondtexture}
\end{figure*}

To investigate the possible valence-bond solid (VBS) order,
we study the dimer-dimer correlation function on cylinder systems,
which is defined as
\begin{eqnarray}
D_{(i,j),(k,l)}=4[\langle(\vec{S}_i \cdot \vec{S}_j)(\vec{S}_k \cdot \vec{S}_l)\rangle-\langle\vec{S}_i \cdot \vec{S}_j\rangle\langle\vec{S}_k \cdot \vec{S}_l\rangle],
\end{eqnarray}
where $(i,j)$ and $(k,l)$ represent the nearest-neighbor (NN) bonds.
First of all, we demonstrate the real space distributions of the NN bond energies on cylinder systems.
To clearly show the fluctuations of the NN bond energies, we define the bond texture as the
bond energies subtracting a constant $e$, which is the average NN bond energy in the bulk of cylinder, i.e.,
$B_{i,j} = \langle S_i\cdot S_j\rangle-e$.
As shown in Suppl. Fig.~\ref{bondtexture} of the bond textures on $3\times 16\times 4$ cylinders for $J^{\prime} = 0.5$
in the vacuum sector, we obtain the
uniform bond textures along both the $x$ and $y$ directions in the bulk of cylinders.
The small differences between the $x$ and $y$ bond textures $0.01$ and $0.004$ in Suppl. Figs. \ref{bondtexture}(a) and \ref{bondtexture}(b)
are owing to the long cylinder geometry, which breaks the lattice rotation symmetry. The uniform bond textures indicate
the good convergence of our DMRG results.

\begin{figure*}
   \centering
   \includegraphics[width=1.0\linewidth]{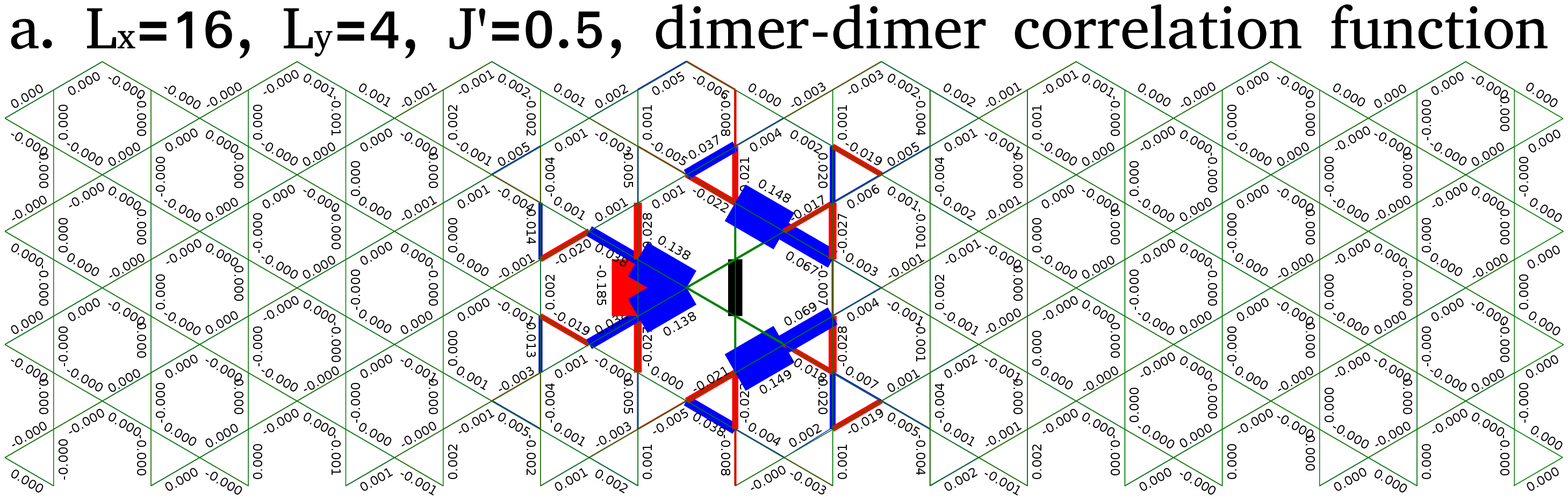}
   \includegraphics[width=1.0\linewidth]{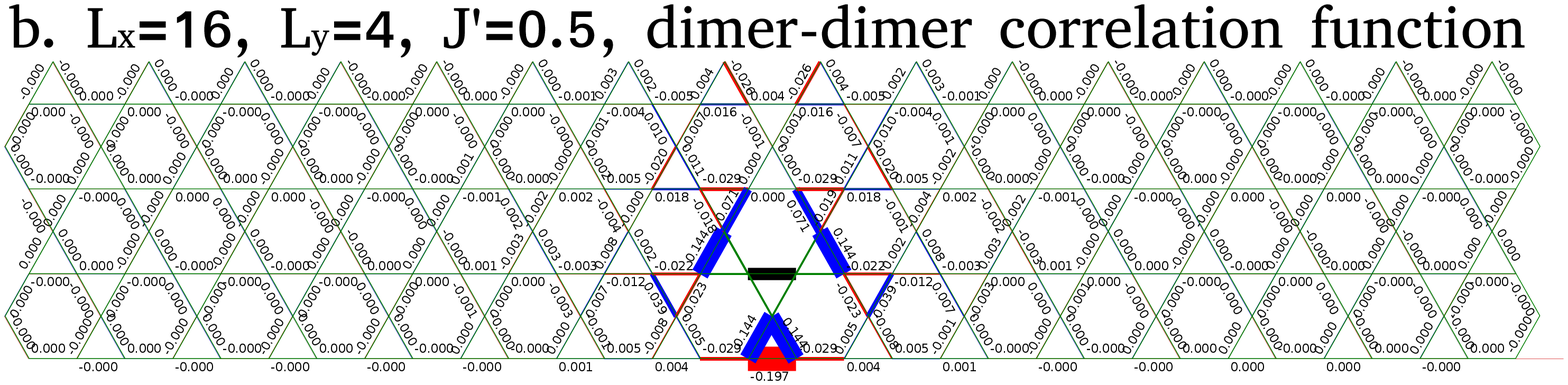}
   \caption{\textbf{Dimer-Dimer correlation function at $J^{\prime}=0.5$}.
   Dimer-Dimer correlation for $J^{\prime}=0.5$ on the $3\times 16\times 4$ cylinders in the vacuum sector.
   The black bond in the middle of cylinder denotes the reference bond $(i,j)$.
   The blue and red bonds represent the positive and negative dimer correlations, respectively.
   }\label{dimerdimer}
\end{figure*}

With the uniform bond textures in the bulk, we could further study the dimer-dimer correlation functions.
We set the reference bond $(i,j)$ in the middle of cylinder.
Suppl. Fig. \ref{dimerdimer} shows the dimer-dimer correlations on the $3\times 16\times 4$ cylinders
at $J^{\prime}=0.5$ in the vacuum sector. The black bond in the middle denotes the reference bond $(i,j)$, and the red and blue bonds indicate the
negative and positive dimer correlations, respectively.
We show that the dimer-dimer correlations decay quite fast to zero in both $x$ and $y$ directions.
On the $3\times 18\times 6$ cylinders, the dimer correlations have a similar fast decay.
The significant short-range dimer correlations strongly indicate the vanishing VBS order.

\begin{figure*}
   \centering
   \includegraphics[width=1.0\linewidth]{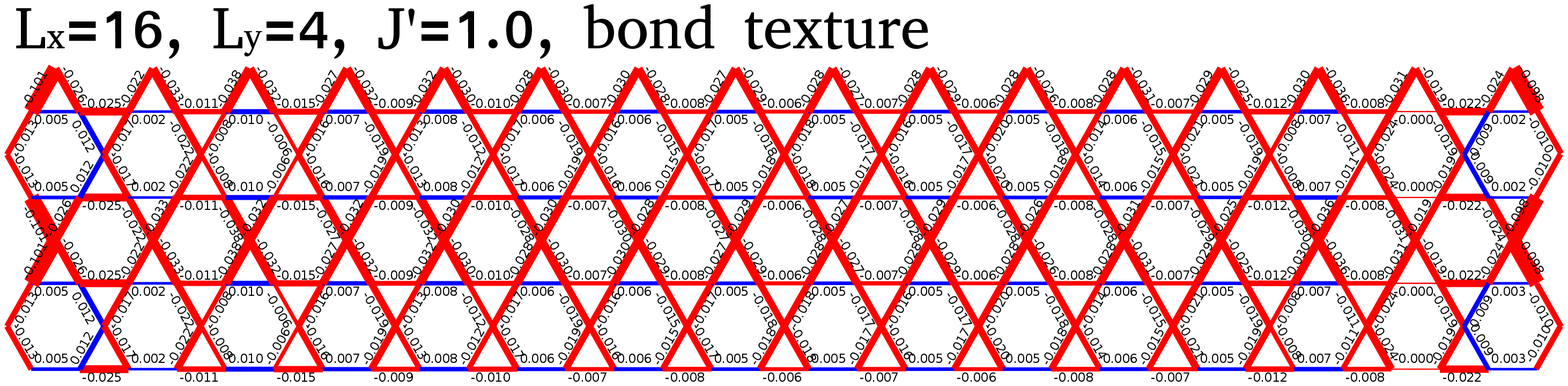}
   \caption{\textbf{Nearest-neighbor bond textures at $J^{\prime} = 1.0$}. The NN bond textures $B_{i,j}$ on the $3\times 16\times 4$ cylinder at
  $J^{\prime} = 1.0$. The numbers denote the amplitudes of bond texture $B_{i,j} = \langle S_i \cdot S_j\rangle-e$,
  where $e$ is the average of the horizontal NN bond energy in the bulk of cylinder. Here, we find $e = -0.0385$.
   The blue (red) color represents the positive (negative) bond texture.}\label{dimertexture}
\end{figure*}

For $J^{\prime}\geq 0.8$, we find a strong VBS state with breaking lattice translational symmetry in the system.
As demonstrated in Suppl. Fig.~\ref{dimertexture} of the bond textures at $J^{\prime}=1.0$ on a $3\times 16\times 4$ cylinder,
the horizontal NN bond textures are not uniform in the bulk of cylinder but have a difference of $0.01$, and the tilt bonds
along the vertical direction also have a difference of $0.01$, which are quite different from the uniform state in the CSL
phase as shown in Suppl. Fig.~\ref{bondtexture}(b). These observations indicate that we find the ground state with lattice
translational symmetry breaking in both the $x$ and $y$ directions.

\section{Chiral-Chiral correlation function}

\begin{figure}
   \centering
   \includegraphics[width=1.0\linewidth]{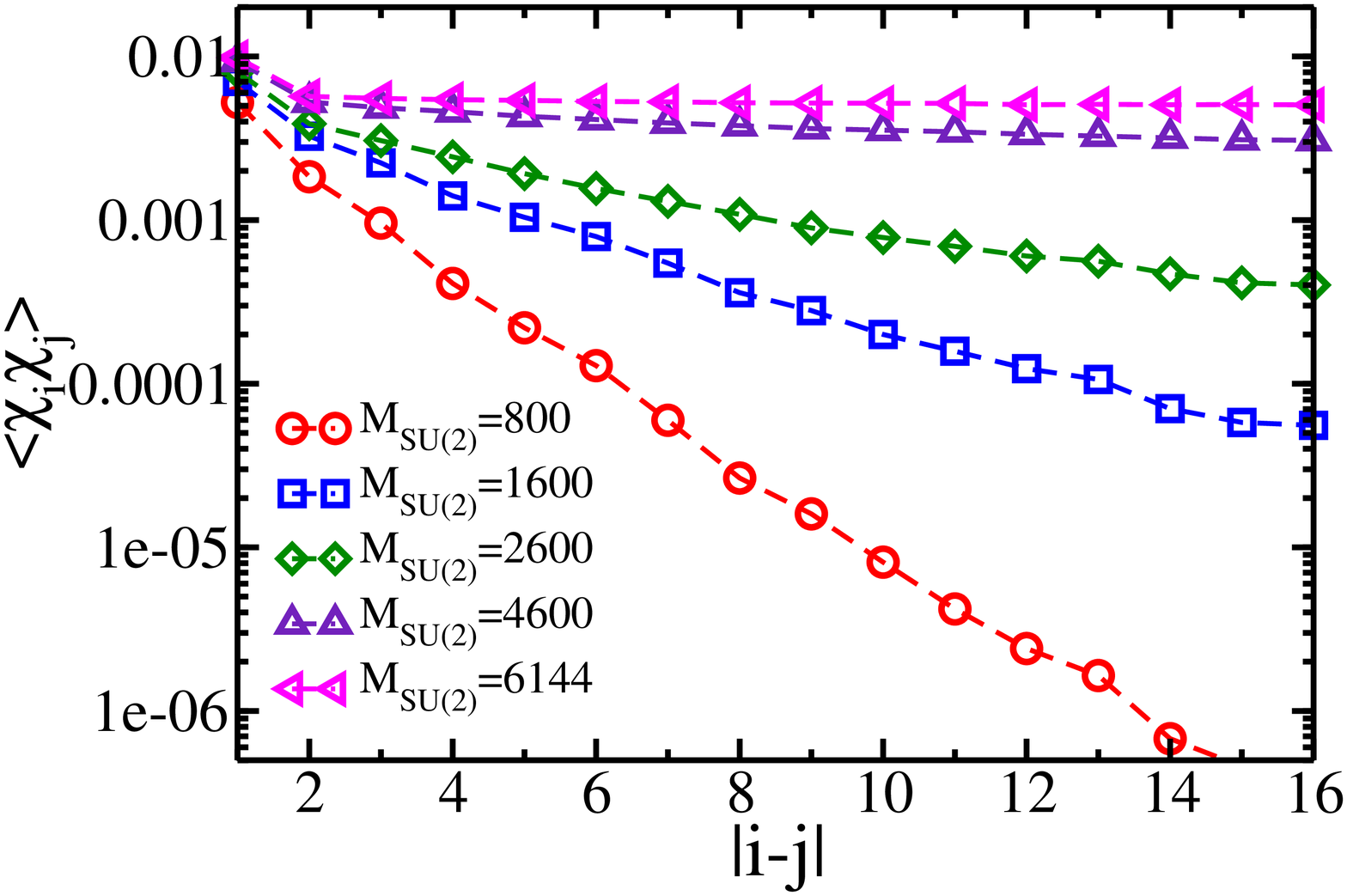}
   \caption{\textbf{Chiral-Chiral correlation function}. The improvement of chiral-chiral correlation function
  with the growing DMRG optimal states for a $3\times 18\times 6$ cylinder at $J^{\prime} = 0.2$ in the vacuum sector.
  $M_{\rm SU(2)}$ is the kept $SU(2)$ states for obtaining the different chiral correlations, which are equivalent to about $3200,6400,10000,18000$, and $24000$ $U(1)$ states.}\label{chiral_M}
\end{figure}

In the DMRG calculations of chiral-chiral correlation function, the systems near phase boundaries require
much more kept states than in deep of the CSL region to capture the long-range chiral correlations.
As shown in Suppl. Fig. \ref{chiral_M} of the improvement of chiral correlation function with the growing
DMRG kept states for a $N = 3\times 18\times 6$ cylinder at $J^{\prime} = 0.2$ in the vacuum sector, the system
shows a fast exponential decay chiral correlation by keeping $800$ $SU(2)$ states (equivalent to about $3200$ $U(1)$ states), and with
increasing kept states the decay length continues to grow. When keeping $4600$ $SU(2)$ states (equivalent to about $18000$ $U(1)$ states), the chiral
correlations form a long-range correlation. Meanwhile, we only need to keep about $10000$ $U(1)$ states to uncover
the long-range chiral correlations in deep of the CSL region such as at $J^{\prime} = 0.4,0.5$.
Therefore, the less convergent DMRG calculations may find a narrower CSL phase region.

\section{Exact Diagonalization results}

\subsection{Lowest-Energy spectrum for 36-sites torus}
\begin{figure}
   \centering
   \includegraphics[width=0.8\linewidth]{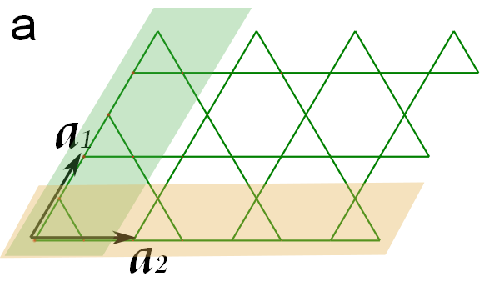}
   \includegraphics[width=0.8\linewidth]{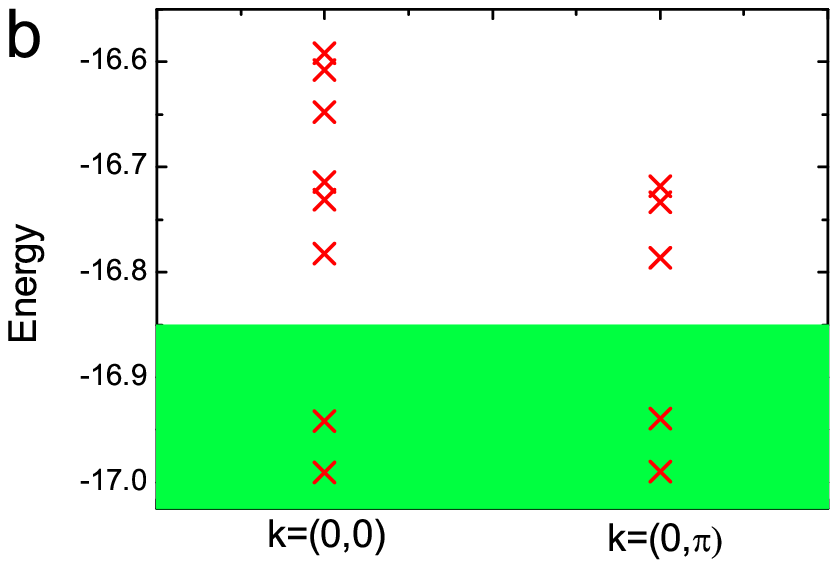}
   \caption{(a) Geometry of the $36$ sites kagome lattice with lattice constant $\vec{a}_1,\vec{a}_2$.
   (b) Low-energy spectrum in the momentum space $\textbf{k}=(0,0),(0,\pi)$ at $J^{\prime}=0.6$.
   }\label{EDenergy}
\end{figure}

We calculate the low-energy spectrum of the $J-J^{\prime}$ model on a $36$ sites kagome lattice
using exact-diagonalization (ED) method. We consider a finite system with periodic boundary conditions, as shown in Suppl. Fig. \ref{EDenergy}(a).
For this geometry, the two-fold topological degeneracy of the $\nu=1/2$ FQHE are expected to live in
the momentum sectors $\mathbf{k}=(0,0)$ and $(0,\pi)$.
Thus we obtain the low-energy spectrum in these two momentum subspaces. As shown in Suppl. Fig. \ref{EDenergy}(b)
of the spectrum at $J^{\prime} = 0.6$,
we find that two lowest states for each momentum sector, denoted by $E^{k=0(\pi)}_1,E^{k=0(\pi)}_2$,
are well separated from the continuum of other excitations by a gap that is about $0.15$.
The nearly vanishing energy difference between two sectors $E^{0}_{1(2)}-E^{\pi}_{1(2)}=0.0007 (0.0022)$ indicates 
the emergence of the many-body magnetic translational symmetry.
The existence of the two lowest states in each sector is due to the time reversal symmetry.
Therefore, our ED calculations imply that the system has four-fold degeneracy of ground states, where
two of them are from topological degeneracy and two are from time reversal symmetry.

\subsection{Modular matrix}
The information of quantum dimension and fusion rules of the quasiparticles are encoded in the
modular $\mathcal{S}$ matrix.
To extract  the modular $\mathcal{S}$ matrix in our model,
we use the method of searching the minimal entropy states (MESs) to construct the modular matrix.
In this method, we first calculate the entanglement entropy
through partitioning the full torus system into two subsystems (cylinders) $A$ and $B$
then tracing out the subsystem $B$.
Here we consider two noncontractible bipartitions on torus geometry as shaded by light green and brown in Suppl. Fig.~\ref{EDenergy}(a),
which is along the lattice vectors $\vec{a}_1,\vec{a}_2$, respectively.

We denote the four groundstates from ED calculation as,
\begin{equation}
  |\psi^{k=0}_1\rangle, |\psi^{k=0}_2\rangle, |\psi^{k=\pi}_1\rangle, |\psi^{k=\pi}_2\rangle.
\end{equation}
Here,  each wavefunction is being chosen as a real one. 
All the above four groundstates preserve the time reversal symmetry and show a vanishing chiral order.
According to the discussions in main text, we can construct the chiral states in each sector as
\begin{eqnarray}
  |\tilde{\psi}^{k=0}_{L,R}\rangle&=& \frac{1}{\sqrt{2}} (|\psi^{k=0}_1\rangle \pm i |\psi^{k=0}_2\rangle) \\
  |\tilde{\psi}^{k=\pi}_{L,R}\rangle&=& \frac{1}{\sqrt{2}} (|\psi^{k=\pi}_1\rangle \pm i |\psi^{k=\pi}_2\rangle)
\end{eqnarray}
where $L(R)$ represents the left (right) chirality.

\begin{figure}
   \centering
   \includegraphics[width=0.8\linewidth]{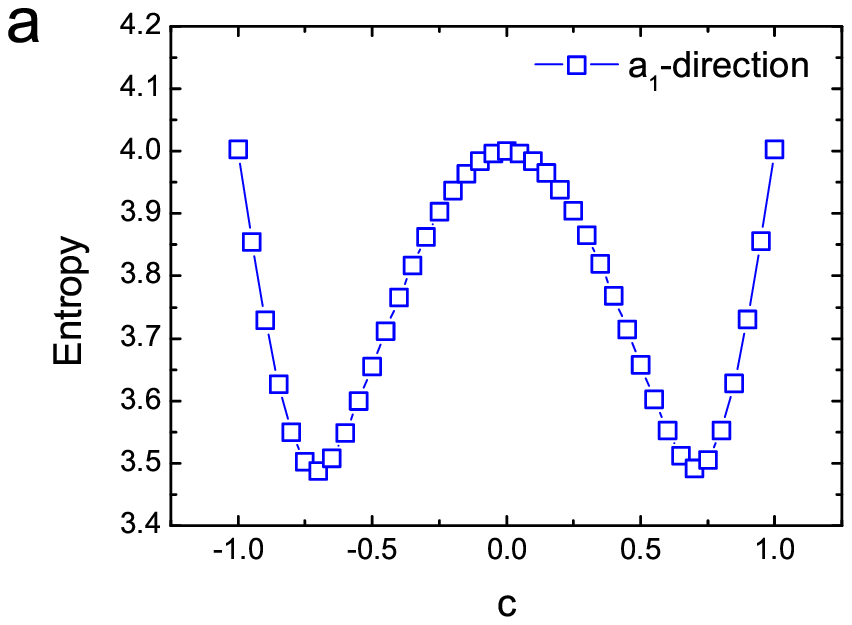}
   \includegraphics[width=0.8\linewidth]{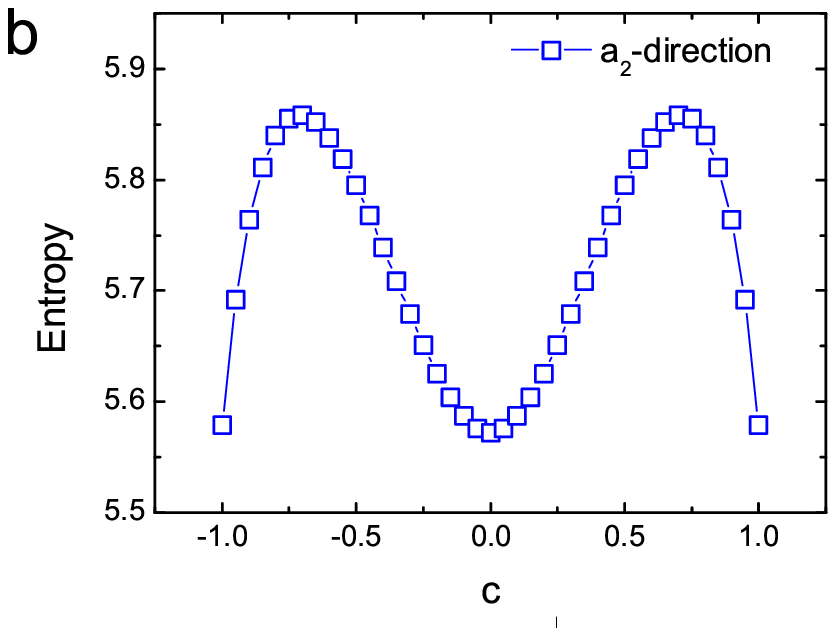}
   \caption{
   Entropy for the superposition state $|\Psi\rangle = c|\tilde{\psi}^{k=0}_{L}\rangle + \sqrt{1-c^2}|\tilde{\psi}^{k=\pi}_{L}\rangle$ for the partition along
   (a) $a_1$-direction and (b) $a_2$-direction. The black arrows show parameters for the MESs. }\label{MES:ED}
\end{figure}

Then we use two chiral states with the same chirality, for example $|\tilde{\psi}^{k=0}_{L}\rangle$ and $|\tilde{\psi}^{k=\pi}_{L}\rangle$,
to calculate the modular matrix.
We search for the MESs in the space of the groundstate manifold
using the following superposition wavefunction:
\begin{equation*}
  |\Psi\rangle = c|\tilde{\psi}^{k=0}_{L}\rangle +\sqrt{1-c^2}e^{i\phi}|\tilde{\psi}^{k=\pi}_{L}\rangle,
\end{equation*}
where $c \in [0,1]$ and $\phi \in [0,2\pi]$ are the superposition parameters.
In our calculation, we find that the global MESs take $\phi = 0$.
As shown in Suppl. Fig. \ref{MES:ED},  the two orthogonal MESs along $a_1$-direction
are respectively located at $c=\frac{1}{\sqrt{2}}$ and $c=-\frac{1}{\sqrt{2}}$,
while the MESs along $a_2$-direction occur at $c=0,1$.
Therefore, we have two MESs along $a_1$-direction
\begin{eqnarray}\label{}
  |\Xi^{a_1}_{1}\rangle&=& \frac{1}{\sqrt{2}} (|\tilde{\psi}^{k=0}_{L}\rangle + |\tilde{\psi}^{k=\pi}_{L}\rangle), \\
  |\Xi^{a_1}_{2}\rangle &=& \frac{1}{\sqrt{2}} (|\tilde{\psi}^{k=0}_{L}\rangle - |\tilde{\psi}^{k=\pi}_{L}\rangle),
\end{eqnarray}
and the two MESs along $a_2$-direction,
\begin{eqnarray}\label{}
  |\Xi^{a_2}_{1}\rangle&=& |\tilde{\psi}^{k=0}_{L}\rangle, \\
  |\Xi^{a_2}_{2}\rangle&=& |\tilde{\psi}^{k=\pi}_{L}\rangle.
\end{eqnarray}

The modular $\mathcal{S}$ matrix is obtained from the overlaps between the MESs of the two noncontractible partition directions:
\begin{equation}
  \mathcal{S}=\langle\Xi^{a_1}|\Xi^{a_2}\rangle=
  \frac{1}{\sqrt{2}}
               \left(
                \begin{array}{cc}
                  1 & 1 \\
                  1 & -1 \\
                \end{array}
              \right),
\end{equation}
which is consistent with the prediction of $SU(2)_1$ conformal field theory about the $\nu=1/2$ bosonic Laughlin state.
Through the modular matrix above, we can extract
the individual quantum dimension $d_{\openone(s)}=1$ for quasiparticle $\openone(s)$
and the fusion rules $\openone\times \openone=\openone$, $\openone\times s=s$, $s\times s=\openone$,
which also determine the characteristic semion statistics of the $s$ quasiparticle.

\end{appendices}

\end{document}